\documentclass{article}
\usepackage[utf8]{inputenc}
\usepackage{geometry}
\newgeometry{hmargin={3cm,3cm}}   
\usepackage{color,amssymb,amsmath,epsfig,ifthen,graphicx,tabularx,xargs}
\usepackage{amsfonts}
\usepackage{amsthm}
\usepackage{algorithm}
\usepackage{algpseudocode}
\usepackage{enumitem,verbatim}
\usepackage{hyperref}
\hypersetup{
    colorlinks,
    linkcolor={red},
    citecolor={blue},
    urlcolor={blue!80!black}
}

\usepackage{cleveref}
\theoremstyle{remark}
\usepackage{mathrsfs}
\usepackage{stmaryrd}
\usepackage{caption}
\usepackage{subcaption}
\usepackage{pdflscape}
\usepackage{authblk}
\usepackage{natbib}
\usepackage{changepage}
\usepackage{booktabs}      
\usepackage{colortbl}      
\usepackage{array}         

\usepackage{xcolor}

\title{\textbf{Competing Risk Analysis in Cardiovascular Outcome Trials: A Simulation Comparison of Cox and Fine-Gray Models}}

\begin{document}

\author{Tuo Wang\thanks{\url{tuo.wang@lilly.com}} and Yu Du\thanks{\url{du_yu@lilly.com}}}
\affil{Global Statistical Sciences, Eli Lilly and Company, Indiana, United States}

\maketitle

\begin{abstract}
Cardiovascular (CV) outcome trials commonly encounter competing risks when non-CV death prevents observation of the primary endpoint, major adverse cardiovascular events (MACE). {Standard Cox proportional hazards models treat competing events as censored, assuming independence between latent event times. Fine-Gray subdistribution hazard models target a different estimand related to cumulative incidence. Despite targeting different quantities,} both approaches are widely used in practice. This simulation study compares the two approaches across clinically relevant scenarios to understand when they yield concordant versus divergent estimates. We conduct Monte Carlo simulations generating correlated time-to-event data using bivariate copula models, systematically varying competing event rates (0.5\% to 5\% annually), treatment effects on the competing event (50\% reduction to 50\% increase in hazard), and correlation between primary and competing events (independent to strong positive). {We derive a closed-form expression for the observed cause-specific hazard ratio under the Gumbel copula and verify it against the simulations.} At competing event rates typical of CV outcome trials (approximately 1\% annually), Cox and Fine-Gray produced nearly identical estimates regardless of correlation strength or treatment effect direction; substantial divergence occurred only when competing rates were high with discordant treatment effects. {The Cox model consistently estimated the observed cause-specific hazard ratio, which differs from the latent treatment effect by an analytically derivable estimand gap that was negligible when competing event rates were low, when primary and competing events were independent, or when treatment effects were concordant.} Pre-specified Cox models should not be abandoned in favor of competing risk methods, and Fine-Gray models do not constitute proper sensitivity analyses to Cox models per ICH E9(R1), as they target different estimands rather than testing assumptions of the same estimand. {The choice between methods should be driven by the scientific estimand of interest, and the estimand gap under high competing-risk scenarios motivates consideration of alternative approaches (inverse probability of censoring weighting, multiple imputation, or inclusion of all-cause mortality in primary composite endpoints) when the marginal causal effect is the target of inference.}

\end{abstract}

\section{Introduction} \label{sec:introduction}

Cardiovascular outcome trials (CVOTs) evaluate treatment effects on composite endpoints that typically include major adverse cardiovascular events (MACE-3: CV death, non-fatal myocardial infarction, or non-fatal stroke). When patients die from non-CV causes before experiencing the primary endpoint, these competing events prevent observation of the primary outcome and create a competing risk scenario requiring careful analytical consideration. In most CVOTs, non-CV death serves as the primary competing risk.

The standard analytical approach in most reported CVOTs uses Cox proportional hazards models \citep{cox1972regression} that treat non-CV death as {censored observations}. Recent large-scale trials with semaglutide illustrate this approach: the SELECT trial evaluates MACE in patients with preexisting CV disease and overweight or obesity but without diabetes \citep{lincoff2023semaglutide}, while the FLOW trial assesses kidney outcomes in patients with type 2 diabetes and chronic kidney disease \citep{perkovic2024effects}. However, {when both CV and non-CV deaths share common risk factors such as age, comorbidity burden, or disease severity, the observed cause-specific hazard ratio estimated by the Cox model may differ from the latent treatment effect on the primary event, a distinction that has important implications for interpretation}.

Journal editors and reviewers, particularly at high-impact venues such as \textit{The New England Journal of Medicine} (NEJM), increasingly scrutinize the handling of competing risks in CV outcome trials. NEJM's statistical reporting guidelines specifically address this issue in their ``Considerations in Time-to-Event Analyses'' \citep{nejm2022guidelines}, stating that ``when a substantial proportion of participants experience a competing event that precludes the occurrence of the primary outcome, alternative analytical approaches such as the Fine-Gray model may be considered.'' The guidelines emphasize evaluating whether censoring by competing events might be informative. However, critical questions remain: What constitutes a ``substantial proportion'' of competing events? At what threshold does the choice between Cox and Fine-Gray models alter trial conclusions? How does the dependency between competing events and primary events impact the result? Is Fine-Gray model the best model to handle competing risk? The NEJM guidelines do not provide quantitative benchmarks for these decisions, leaving investigators uncertain about when competing risk methods add value versus unnecessary complexity.

When competing events are present, two analytical approaches are commonly used, each targeting fundamentally different estimands. Table~\ref{tab:model_comparison} summarizes key distinctions between the Cox proportional hazards model and the Fine-Gray subdistribution hazard model. Despite targeting different estimands, both approaches are widely used in practice, and understanding when they yield concordant versus divergent numerical results has important practical implications for trial design, analysis planning, and interpretation. A key question is whether the choice between these methods affects trial conclusions in typical CVOT settings.

\begin{table}[htbp]
\centering
\caption{Comparison of Cox Proportional Hazards and Fine-Gray Subdistribution Hazard Models in the Presence of Competing Risks}
\label{tab:model_comparison}
\small
\begin{tabular}{>{\raggedright\arraybackslash}p{3.5cm}>{\raggedright\arraybackslash}p{5.5cm}>{\raggedright\arraybackslash}p{5.5cm}}
\toprule
\textbf{Dimension} & \textbf{Cox Proportional Hazards Model} & \textbf{Fine-Gray Subdistribution Hazard Model} \\
\midrule
\rowcolor{gray!10}
\textbf{Target estimand} 
& Cause-specific hazard ratio 
& Subdistribution hazard ratio \\
\addlinespace[0.3em]
\textbf{Risk set definition} 
& Individuals who are event-free at each time point (competing events treated as censored)
& Individuals who are event-free, including those who have experienced a competing event (who remain in the risk set with time-dependent weights) \\
\addlinespace[0.3em]
\rowcolor{gray!10}
\textbf{Handling of competing events}
& {Competing events are treated as censored (computational device for cause-specific hazard estimation)}
& Competing events are explicitly incorporated through weighting in the subdistribution hazard \\
\addlinespace[0.3em]
\textbf{Key model assumption}
& {Proportional cause-specific hazards; non-informative administrative censoring}
& Proportional subdistribution hazards \\
\addlinespace[0.3em]
\rowcolor{gray!10}
\textbf{Interpretation of treatment effect} 
& Treatment effect on the instantaneous rate of the primary event among individuals who remain event-free
& Treatment effect on the instantaneous rate of the primary event in the subdistribution (including those event-free or having experienced competing events) \\
\bottomrule
\end{tabular}
\end{table}


\textbf{The Cox proportional hazards model} \citep{cox1972regression} is the most widely used approach in CVOTs, treating competing events as censored observations. This model estimates the instantaneous risk (hazard) of the primary CV event among patients still at risk. The hazard ratio quantifies the treatment effect on the rate of primary event occurrence, conditional on not yet having experienced either the primary or competing event. This estimand addresses: ``Among patients who remain event-free, how does treatment affect the rate of primary CV events?'' This interpretation is straightforward and aligns naturally with clinical reasoning about treatment mechanisms.

In the SOUL trial, the primary Cox analysis yields a hazard ratio of 0.86 (95\% CI: 0.77-0.96) for MACE-3 \citep{mcguire2025oral}, with non-CV death occurring at approximately 1.4 per 100 patient-years.


\textbf{The Fine-Gray subdistribution hazard model} \citep{fine1999proportional} is often recommended for explicitly handling competing risks. This approach models the subdistribution hazard, which relates directly to cumulative incidence functions. The subdistribution hazard ratio (sHR) quantifies treatment effects on the subdistribution hazard, the instantaneous rate of primary event occurrence in the subdistribution, which includes both patients who remain at risk and those who have experienced competing events (with modified weights). While the subdistribution hazard relates to cumulative incidence probabilities, the subdistribution hazard itself lacks a direct clinical interpretation as an instantaneous rate for any identifiable group of patients \citep{austin2017practical,austin2021fine,armbruster2024pitfalls,gregson2024competing}.

As an empirical example, the FIGARO-DKD trial \citep{pitt2021cardiovascular} evaluated the effect of finerenone on CV outcomes in patients with chronic kidney disease and type 2 diabetes. The primary Cox analysis yielded a hazard ratio of 0.87 (95\% CI: 0.76-0.98) for the primary composite CV outcome, while the Fine-Gray competing risk analysis accounting for death as a competing event yielded a subdistribution hazard ratio of 0.87 (95\% CI: 0.77-0.98), demonstrating numerical concordance between methods. Notably, the finerenone FDA label includes only Cox model results, not Fine-Gray model results, highlighting regulatory acceptance of Cox models as the primary analytical approach when competing event rates are low.

A critical but often overlooked issue is that {the Cox model consistently estimates the cause-specific hazard ratio, which generally differs from the latent marginal treatment effect parameter when competing risks are dependent and treatment affects both event types. Understanding when this estimand gap is negligible versus substantial has important implications for method selection.} Despite extensive methodological literature on competing risks, practical guidance remains limited regarding when the choice between Cox and Fine-Gray models affects trial conclusions in CVOTs. {This simulation study addresses these gaps by systematically comparing Cox proportional hazards models and Fine-Gray subdistribution hazard models across parameter ranges relevant to CV outcome trials, deriving a closed-form expression for the observed cause-specific hazard ratio under the copula model, and verifying it against Monte Carlo simulations.}

{Prior simulation work in this area includes \citet{beyersmann2009simulating}, who established best practices for generating competing risks data via cause-specific hazards, and \citet{latouche2007misspecified}, who examined the behavior of the Fine-Gray model under misspecification. Our study extends this literature by using a copula-based DGM that explicitly parameterizes and varies the dependence between event types, a dimension absent from earlier work. We also calibrate all parameters (event rates, treatment effects, follow-up duration) to contemporary CVOT data rather than abstract settings, and we systematically map the region of parameter space where the two methods diverge, providing quantitative thresholds rather than qualitative guidance.}

Specifically, we aim to: (1) quantify the divergence between Cox and Fine-Gray model estimates across realistic scenarios; (2) identify parameter thresholds where model choice substantially impacts trial interpretation; (3) characterize scenarios where neither estimator provides consistent recovery of the marginal hazard ratio; and (4) provide evidence-based guidance for analytical approach selection in CVOTs.

Section~\ref{sec:methods} describes the simulation DGM and statistical analyses. Section~\ref{sec:results} presents the primary comparison and robustness checks. Section~\ref{sec:empirical} provides an empirical illustration from a published trial. Section~\ref{sec:discussion} discusses practical implications, and study limitations. Section~\ref{sec:conclusion} offers conclusions.

\section{Methods} \label{sec:methods}

We conduct a Monte Carlo simulation study {with parameters calibrated to contemporary CV outcome trials}, where MACE-3 serves as the primary endpoint and non-CV death acts as the competing risk. 

\subsection{Simulation Set-up}

We generate correlated event times using a bivariate Gumbel-Hougaard copula, which allows flexible specification of dependence between primary CV events and competing non-CV death while maintaining exponential marginal distributions {\citep[see][for general principles of competing risks simulation]{beyersmann2009simulating}}. Let $T_1$ denote the time to first MACE-3 event and $T_2$ denote time to competing non-CV death. For a patient with treatment assignment $Z$ (where $Z=1$ indicates treatment and $Z=0$ indicates control), the joint survival function is:

\begin{equation}
S(t_1, t_2 | Z = z) = P(T_1 > t_1, T_2 > t_2 | Z = z) = \exp\{-[(\theta_1^z \cdot \lambda_1 \cdot t_1)^\alpha + (\theta_2^z \cdot \lambda_2 \cdot t_2)^\alpha]^{1/\alpha}\}
\end{equation}

where $\lambda_1$ and $\lambda_2$ are baseline event rates, $\theta_1$ and $\theta_2$ are {the latent treatment effect parameters for each event type (i.e., hazard ratios governing the marginal distributions of $T_1$ and $T_2$ within the copula model)}, and $\alpha$ is the copula dependence parameter that controls the correlation between events \citep{oakes1989bivariate}.

\subsubsection*{Three Estimands: Latent Treatment Effect, Observed Cause-Specific HR, and Subdistribution HR}

We distinguish three quantities central to the interpretation of this simulation study.

\textbf{(i) The latent treatment effect parameter $\theta_1$.} Under the copula DGM, the marginal survival for $T_1$ given treatment $Z$ is $S_1(t \mid Z) = \exp(-\lambda_1 \theta_1^Z \cdot t)$, so the marginal cause-specific hazard is $h_1^{\text{marg}}(t \mid Z) = \lambda_1 \theta_1^Z$ and the marginal hazard ratio is exactly $\theta_1$. This is the treatment's \textit{latent} effect on the primary event, defined within the model prior to any risk-set depletion by competing events. Importantly, $\theta_1$ is a model-dependent quantity: it is known by construction in the simulation but is not generally identifiable from observed competing-risk data $(Y_i, \delta_i)$ without the copula specification, because the joint distribution of the latent event times $(T_1, T_2)$ is not identified from the observed minimum and event indicator under dependent competing risks \citep{tsiatis1975nonidentifiability}. It is neither a causal ``elimination of competing events'' estimand in the sense of \citet{young2020causal} nor a conditional hazard ratio among survivors.

\textbf{(ii) The observed cause-specific hazard ratio $\text{HR}_\text{CS}$.} In a competing-risk dataset, one observes $Y = \min(T_1, T_2, C)$ and $\delta = \{1, 2, 0\}$. The \textit{observed} cause-specific hazard for event 1, the instantaneous rate of the primary event among subjects still free of \textit{all} events at time $t$, is:
\begin{equation}
h_1^{\text{cs}}(t \mid Z) = \lim_{\Delta t \to 0} \frac{P(t \leq T_1 < t + \Delta t,\; T_1 \leq T_2 \mid T_1 \geq t,\; T_2 \geq t,\; Z)}{\Delta t}
\end{equation}
Under the Gumbel-Hougaard copula in Equation~(1), this evaluates to (see Appendix A for derivation):
\begin{equation}
h_1^{\text{cs}}(t \mid Z) = (\lambda_1 \theta_1^Z)^\alpha \cdot \bigl[(\lambda_1 \theta_1^Z)^\alpha + (\lambda_2 \theta_2^Z)^\alpha\bigr]^{1/\alpha - 1}
\end{equation}
Because both marginal hazards are constant (exponential distributions), $h_1^{\text{cs}}(t \mid Z)$ is also constant in time, and the observed cause-specific hazard ratio is:
\begin{equation}
\text{HR}_{\text{CS}} = \frac{h_1^{\text{cs}}(t \mid Z=1)}{h_1^{\text{cs}}(t \mid Z=0)} = \theta_1^\alpha \cdot \left[\frac{(\lambda_1 \theta_1)^\alpha + (\lambda_2 \theta_2)^\alpha}{\lambda_1^\alpha + \lambda_2^\alpha}\right]^{1/\alpha - 1}
\end{equation}
This is the quantity the Cox model \textit{consistently} estimates when it treats competing events as censored. It equals $\theta_1$ under two special conditions: (a) $\alpha = 1$ (independence), or (b) $\theta_2 = \theta_1$ (concordant treatment effects). Outside these conditions, $\text{HR}_{\text{CS}} \neq \theta_1$, not because the Cox estimator fails, but because the observed cause-specific hazard is shaped by the competing-risk mechanism.

\textbf{(iii) The subdistribution hazard ratio.} The Fine-Gray model targets the subdistribution hazard, which conditions on not having experienced the \textit{primary} event (retaining competing-event subjects with redistribution weights). Under the copula DGM, proportional subdistribution hazards do not generally hold, so the Fine-Gray coefficient estimates a time-averaged summary of the subdistribution hazard ratio over the follow-up period. We denote this summary $\text{sHR}_{\text{FG}}$.

\textbf{The estimand gap.} We define the \textit{estimand gap} as $\text{HR}_{\text{CS}} - \theta_1$, the difference between what Cox estimates and the latent treatment effect. This gap is fully determined by Equation~(4) and vanishes when $\alpha = 1$ or $\theta_2 = \theta_1$. In Section~\ref{sec:results}, we verify that the closed-form expression in Equation~(4) predicts the simulation results, and we characterize the parameter regions where the gap is negligible versus substantial. We also examine the analogous gap $\text{sHR}_{\text{FG}} - \theta_1$ for the Fine-Gray model.

The reason $\theta_1$ remains scientifically relevant despite its latent nature is that it represents the treatment's direct pharmacological effect on the primary event. When investigators design a CVOT and power it based on historical Cox estimates, they implicitly assume $\text{HR}_{\text{CS}} \approx \theta_1$, an assumption that holds when competing rates are low (making the term in brackets in Equation~(4) close to 1). The estimand gap quantifies how much the competing-risk mechanism distorts what Cox observes relative to this intended treatment effect.

\textbf{Data generation steps.} For each simulated trial dataset:
\begin{enumerate}
\item \textbf{Randomization:} Generate treatment assignment $Z_i \sim \text{Binomial}(n = 1, p = 0.5)$ for each patient

\item \textbf{Correlated uniform variables via Gumbel-Hougaard copula:} Generate bivariate uniform random variables $(U_{1i}, U_{2i})$ that follow the Gumbel-Hougaard copula structure. The copula data generation is implemented using the \texttt{gumbel} package in R \citep{gumbel2024package}. 

The parameter $\alpha$ determines correlation between $T_1$ and $T_2$ through Kendall's tau: $\tau = 1 - 1/\alpha$. We examine four correlation levels representing hypothesized varying degrees of shared risk factors and pathophysiology:
\begin{itemize}
\item $\alpha = 1.0$ ($\tau = 0$): Independence
\item $\alpha = 1.2$ ($\tau = 0.167$): Weak positive correlation
\item $\alpha = 1.5$ ($\tau = 0.333$): Moderate positive correlation
\item $\alpha = 2.0$ ($\tau = 0.500$): Strong positive correlation
\end{itemize}

\item \textbf{Transform to event times:} Convert correlated uniform variables to event times based on inverse of the cumulative distribution function:
\begin{align*}
T_{1i} &= -\log(U_{1i}) / (\lambda_1 \cdot \theta_1^{Z_i}) \\
T_{2i} &= -\log(U_{2i}) / (\lambda_2 \cdot \theta_2^{Z_i})
\end{align*}
\item \textbf{Administrative censoring:} Generate $C_i \sim \text{Uniform}(3, 5)$ years to mimic staggered uniform enrollment of 2 years and maximum follow-up of 5 years. 

\item \textbf{Determine observed outcomes:}
\begin{itemize}
\item If $T_{1i} < \min(T_{2i}, C_i)$: Observe primary CV event at time $T_{1i}$ 
\item Else if $T_{2i} < C_i$: Observe competing non-CV death at time $T_{2i}$ 
\item Else: Right-censored at time $C_i$ 
\end{itemize}
\end{enumerate}

For each simulated trial, we randomized $n = 500$ patients 1:1 to treatment or control, generated correlated event times from the copula with $\lambda_1 = 0.035$ per patient-year and $\theta_1 = 0.80$, DGM input values chosen so that the implied primary-event incidence (approximately 3.5\% annually in the control arm) and marginal rate ratio are of the same order as contemporary CVOT Cox estimates; {because $\theta_1$ is a latent DGM parameter rather than an estimable quantity (see Section~2.1), this calibration sets a plausible working value rather than recovering $\theta_1$ from historical data}, applied administrative censoring uniformly distributed over 3 to 5 years, and replicated each scenario 2,000 times. We selected a sample size of $n = 500$ to maintain computational efficiency. {Sensitivity analyses at larger sample sizes confirmed that the numerical difference patterns were invariant to $n$ (Section~3.3, Figure~S1).}

We systematically varied three parameters across scenarios. The copula dependence parameter $\alpha$ takes values of 1.0, 1.2, 1.5, and 2.0, corresponding to Kendall's $\tau$ of 0 (independence), 0.167 (weak correlation), 0.333 (moderate correlation), and 0.500 (strong correlation), representing varying degrees of shared risk factors between CV and non-CV events. The competing event rate $\lambda_2$ ranges from 0.005 to 0.05 per patient-year (corresponding to annual hazard rates of 0.5\% to 5\%; at these low rates, annual hazards and annual event probabilities are approximately equal), spanning from very low to moderately high competing risk, with historical CVOTs suggesting typical rates of approximately 1\% annually for non-CV death \citep{lincoff2023semaglutide,mcguire2025oral}. The treatment effect on competing events $\theta_2$ ranges from 0.5 to 1.5, representing three clinically relevant scenarios:

\begin{itemize}
\item $\theta_2 < 1$: Treatment reduces competing events (directionally consistent benefits)
\item $\theta_2 = 1$: Treatment has no effect on competing events (neutral scenario)
\item $\theta_2 > 1$: Treatment increases competing events (directionally discordant effects)
\end{itemize}

\subsection{Statistical Analyses}

For each simulated dataset, we fit two models:

\textbf{Cox proportional hazards model treating competing events as censored.} The Cox model estimates the hazard ratio for the primary CV event, treating non-CV deaths as censored observations. The model for the hazard function is:

\begin{equation}
h(t | Z) = h_{0}(t) \cdot \exp(\beta_{\text{Cox}} Z)
\end{equation}

where $h_{0}(t)$ is the baseline hazard for the primary event and the treatment effect is quantified by $\widehat{\text{HR}}_{\text{Cox}} = \exp(\beta_{\text{Cox}})$. This hazard ratio represents the treatment effect on the instantaneous rate of primary CV events among patients who remain event-free, conditional on not yet having experienced either the primary or competing event. This estimand addresses: ``Among patients who remain event-free, how does treatment affect the rate of primary CV events?'' The Cox model is implemented using the \texttt{survival} package in R \citep{therneau2023survival}.

\textbf{Fine-Gray subdistribution hazard model.} The Fine-Gray model estimates the subdistribution hazard ratio, which relates to cumulative incidence. The model for the subdistribution hazard function is:

\begin{equation}
h^*(t | Z) = h_{0}^*(t) \cdot \exp(\beta_{\text{FG}}  Z)
\end{equation}

where $h_{0}^*(t)$ is the baseline subdistribution hazard and the treatment effect is quantified by $\widehat{\text{sHR}}_{\text{FG}} = \exp(\beta_{\text{FG}})$. The subdistribution hazard ratio quantifies treatment effects on the subdistribution hazard, the instantaneous rate of the primary event occurrence in the subdistribution, which includes both patients who remain at risk and those who have experienced competing events (with modified weights). While the subdistribution hazard itself lacks a direct clinical interpretation as an instantaneous rate, it directly corresponds to treatment effects on cumulative incidence probabilities. The Fine-Gray model is implemented using the \texttt{cmprsk} package in R \citep{gray2022cmprsk}. {For each scenario, we report the exponentiated mean of the log hazard ratios across replications, $\exp(\bar{\beta})$, as the summary point estimate for both Cox and Fine-Gray models, along with the standard deviation of the log hazard ratios as the Monte Carlo standard error.}

\subsubsection*{Estimand Architecture}

The three estimands relate as follows. The Cox model consistently estimates $\text{HR}_{\text{CS}}$ (Equation~4), which is the observed cause-specific hazard ratio, a well-defined, identifiable quantity that characterizes treatment effects on the primary event rate among subjects currently free of all events. The Fine-Gray model estimates a time-averaged summary of the subdistribution hazard ratio, characterizing treatment effects on the cumulative incidence of the primary event in the presence of competing risks. Neither model targets the latent $\theta_1$ directly.

The distinction between $\text{HR}_{\text{CS}}$ and $\theta_1$ arises because treatment differentially depletes the shared risk set. When $\theta_2 \neq \theta_1$, the competing event removes patients at different rates across treatment arms, altering the composition of the at-risk population at each time point. This selection produces an observed cause-specific hazard ratio that differs from the latent marginal rate ratio $\theta_1$. The estimand gap is zero when the selection is symmetric ($\theta_2 = \theta_1$) or absent ($\alpha = 1$).

We note that treating competing events as censored in the Cox model is the standard computational device for estimating the cause-specific hazard; it does \textit{not} require or invoke independence between the latent event times $T_1$ and $T_2$. The cause-specific hazard $h_1^{\text{cs}}(t)$ is defined entirely in terms of observable quantities ($Y$, $\delta$) and is consistently estimated by the Nelson-Aalen/Cox approach regardless of the dependence structure between latent times \citep{andersen2012competing}. Independence assumptions become relevant only when the inferential target is the latent marginal hazard ($\theta_1$) or a hypothetical elimination estimand, estimands that require modeling the unobserved joint distribution.

The estimand distinction also clarifies why comparing Cox and Fine-Gray results cannot serve as a sensitivity analysis per ICH~E9(R1). Discrepancies between $\text{HR}_{\text{Cox}}$ and $\text{sHR}_{\text{FG}}$ reflect that the two models address different scientific questions, not that either result is fragile \citep{andersen2012competing,young2020causal}.

\section{Results} \label{sec:results}

Across all simulation scenarios including wide ranges of competing event rates ($\lambda_2$ from 0.5\% to 5\% annually), treatment effects on competing events ($\theta_2$ from 0.5 to 1.5), and event correlations (Kendall's $\tau$ from 0 to 0.50), we found that Cox and Fine-Gray methods generally produced concordant results. In typical CVOT settings with non-CV death rates of approximately 1\% annually, as observed in trials such as SELECT and SOUL, the two methods produced very similar estimates on average. This concordance held across varying degrees of correlation between CV and non-CV events and across different treatment effect profiles on competing events, provided the competing event rate remained low.

The divergence between Cox and Fine-Gray estimates was primarily driven by three factors: the competing event rate, the treatment effect on competing events, and the correlation between primary and competing events. We examined the impact of each factor while holding the others constant to understand their individual contributions to method divergence.

\subsection{Comparison Between Cox and Fine-Gray Models}

When events were independent ($\alpha = 1.0$, $\tau = 0$), Cox and Fine-Gray models yielded nearly identical estimates. As event correlation increased, Fine-Gray subdistribution hazard ratios tended to diverge from Cox estimates, but even at strong correlation ($\alpha = 2.0$, $\tau = 0.50$), the mean absolute difference remained modest.

\begin{figure}[ht]
\centering
\includegraphics[width=\textwidth]{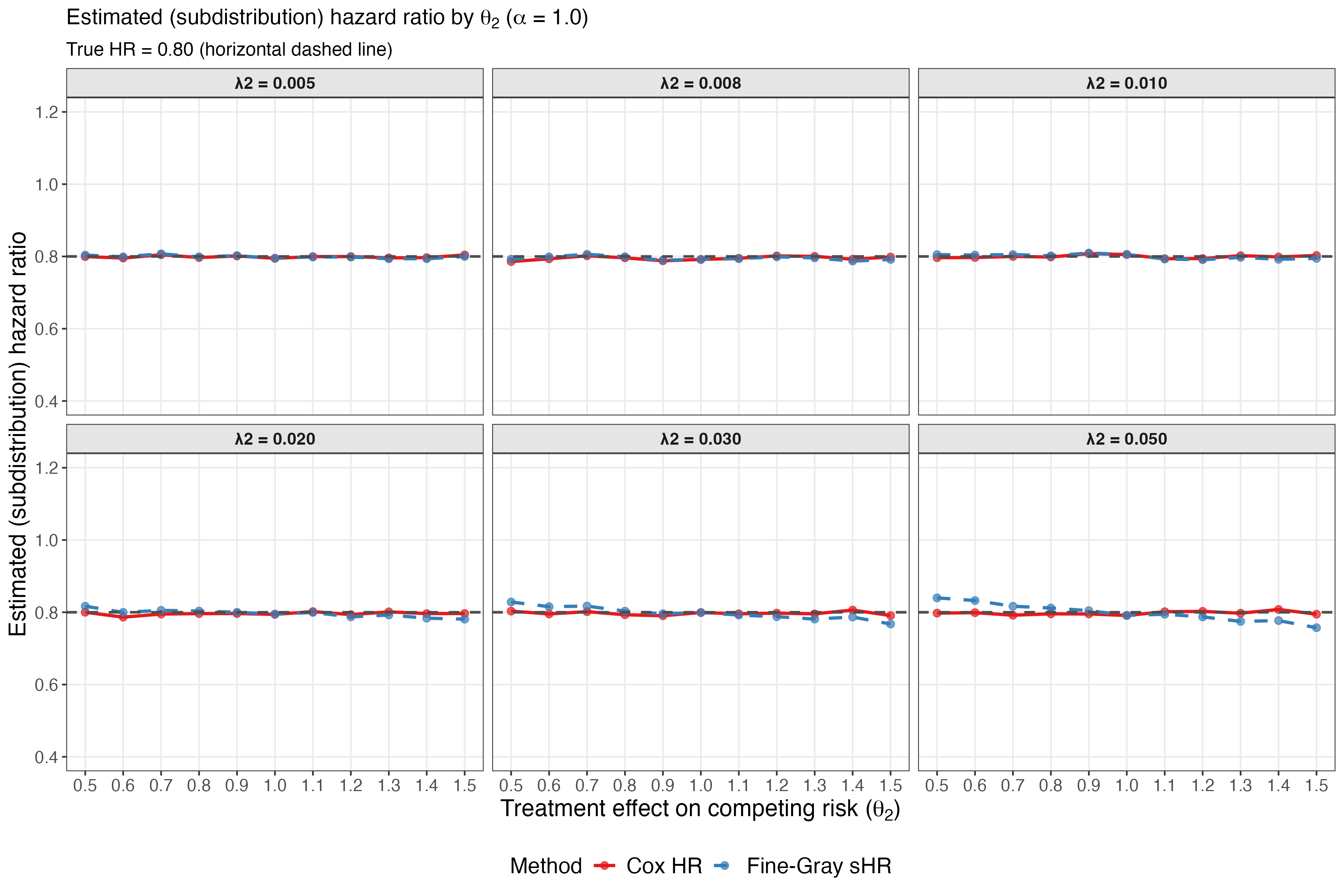}
\caption{{Estimated hazard ratios by treatment effect on competing risk ($\theta_2$) under independence ($\alpha = 1.0$, Kendall's $\tau = 0$). Each panel shows a different competing event rate ($\lambda_2$); red solid line = Cox cause-specific HR, blue dashed line = Fine-Gray subdistribution HR. Horizontal dashed line at 0.80 marks the latent treatment effect parameter $\theta_1$. Under independence ($\alpha = 1$), Equation~(4) gives $\text{HR}_{\text{CS}} = \theta_1$, and both estimators closely approximate $\theta_1$ across all $\lambda_2$ and $\theta_2$ values.}}
\label{fig:lineplot_alpha1}
\end{figure}

\begin{figure}[ht]
\centering
\includegraphics[width=\textwidth]{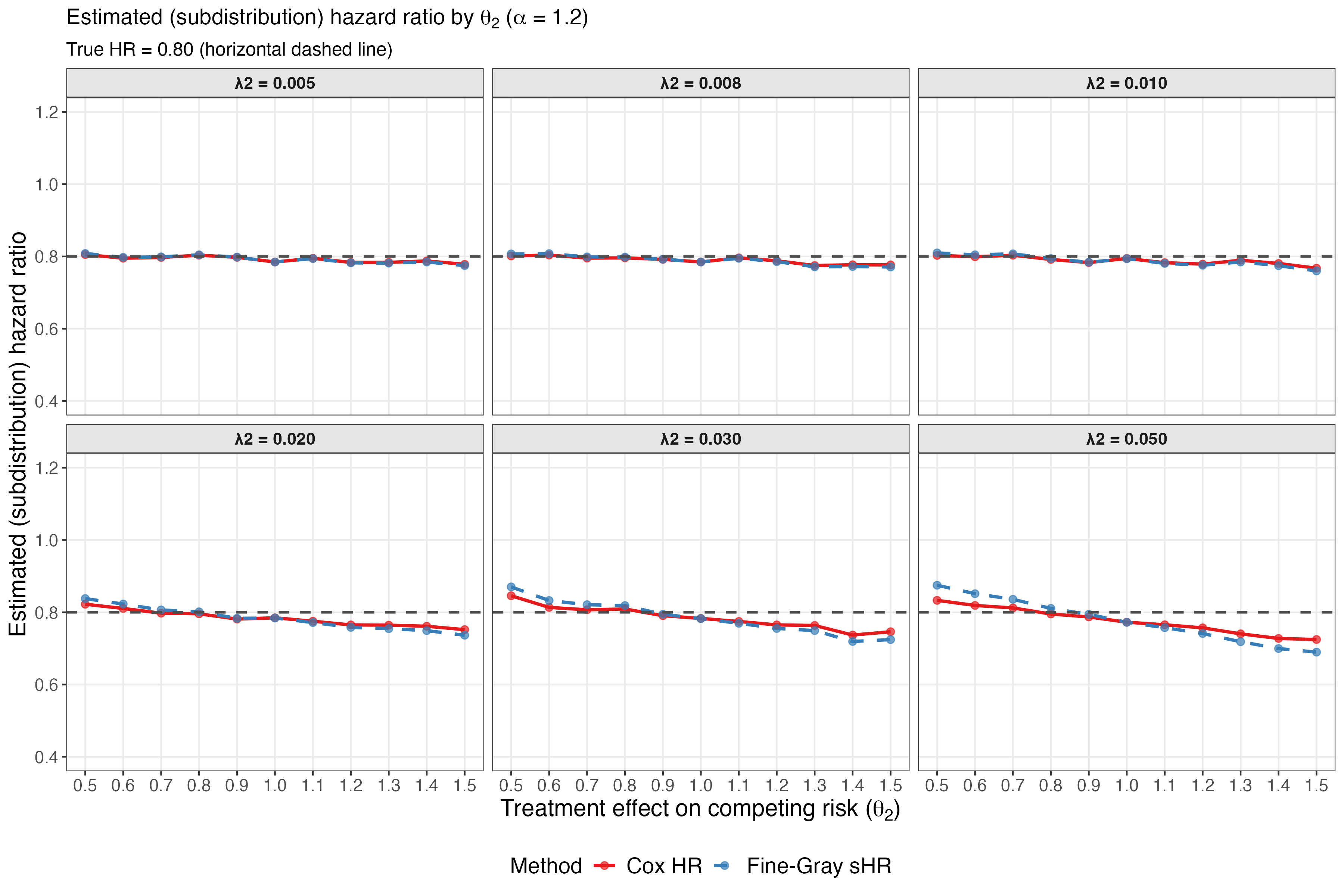}
\caption{{Estimated hazard ratios by $\theta_2$ under weak positive correlation ($\alpha = 1.2$, Kendall's $\tau = 0.167$). Format as in Figure~\ref{fig:lineplot_alpha1}. Weak correlation introduces slight divergence between methods at high $\lambda_2$ and extreme $\theta_2$, but both estimators remain close to $\theta_1 = 0.80$ at low competing rates ($\lambda_2 \leq 0.01$).}}
\label{fig:lineplot_alpha12}
\end{figure}

\begin{figure}[ht]
\centering
\includegraphics[width=\textwidth]{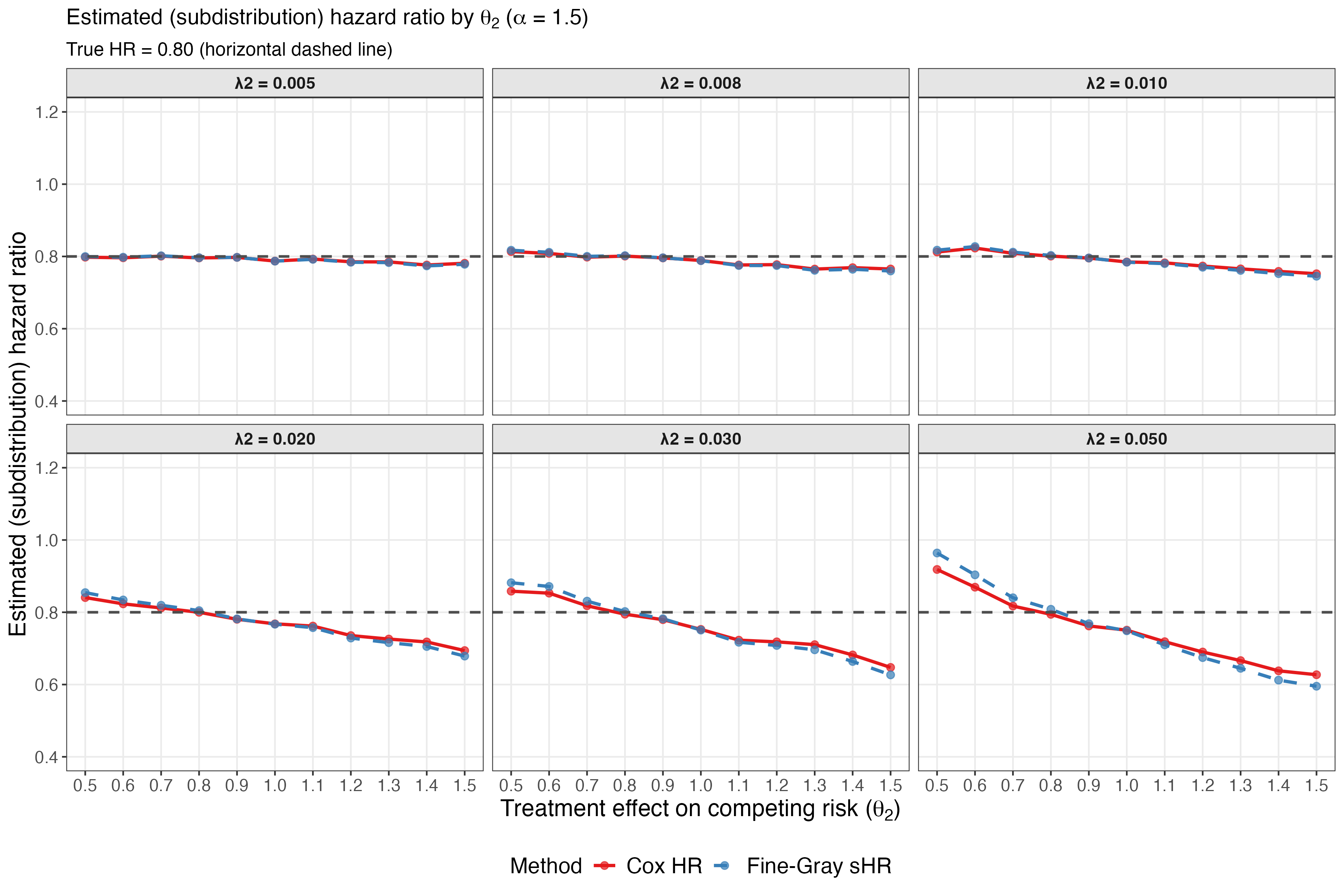}
\caption{{Estimated hazard ratios by $\theta_2$ under moderate positive correlation ($\alpha = 1.5$, Kendall's $\tau = 0.333$). Format as in Figure~\ref{fig:lineplot_alpha1}. Method divergence grows with $\lambda_2$; at $\lambda_2 = 0.05$, the Fine-Gray sHR deviates further from 0.80 than the Cox HR when $\theta_2 < 0.80$ and closer when $\theta_2 > 0.80$.}}
\label{fig:lineplot_alpha15}
\end{figure}

\begin{figure}[ht]
\centering
\includegraphics[width=\textwidth]{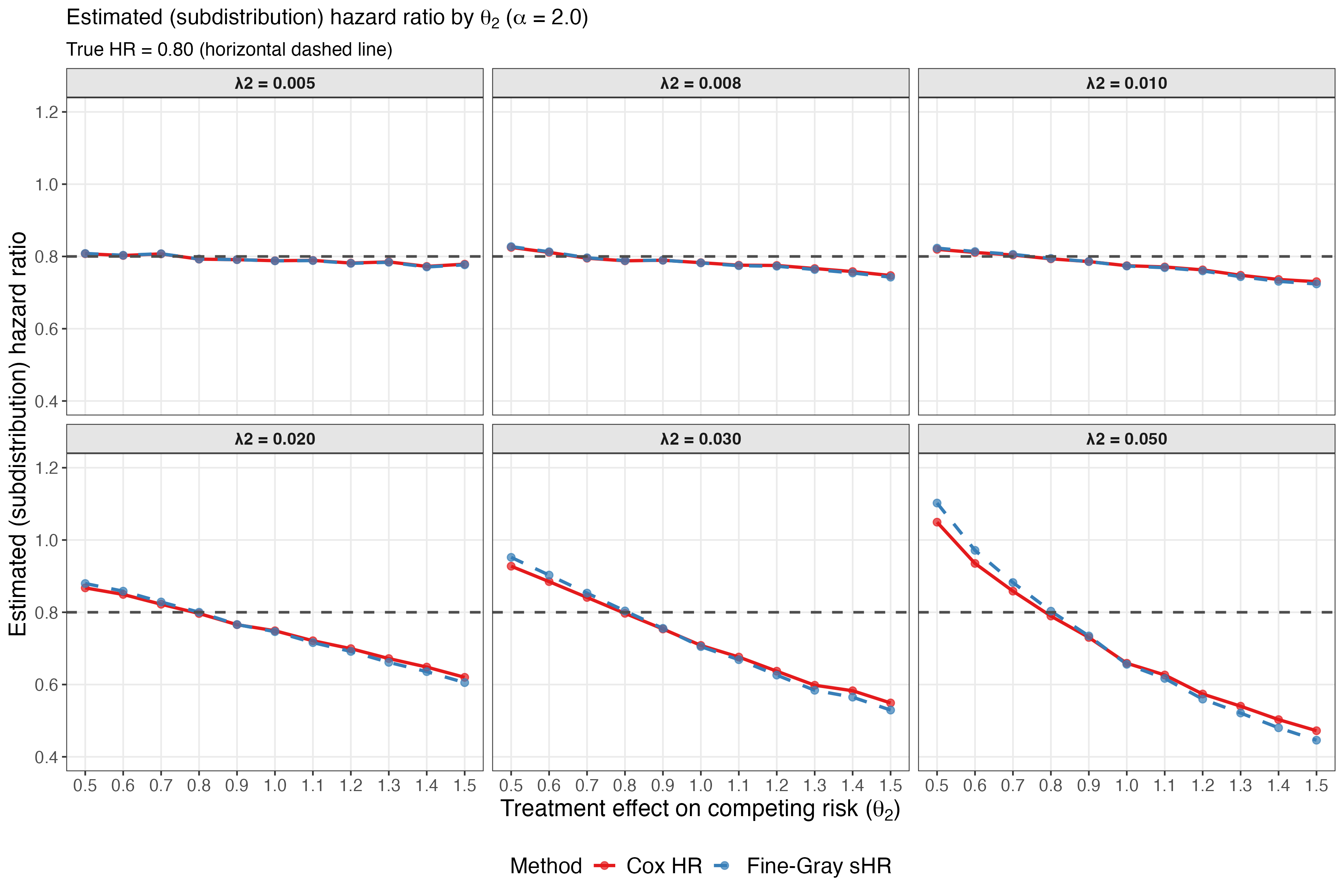}
\caption{{Estimated hazard ratios by $\theta_2$ under strong positive correlation ($\alpha = 2.0$, Kendall's $\tau = 0.50$). Format as in Figure~\ref{fig:lineplot_alpha1}. Strong correlation produces the widest range of deviation from $\theta_1 = 0.80$: at $\lambda_2 = 0.05$, Cox HR ranges from approximately 0.47 ($\theta_2 = 1.5$) to 1.05 ($\theta_2 = 0.5$). Even here, concordance between methods at low competing rates ($\lambda_2 \leq 0.01$) remains high.}}
\label{fig:lineplot_alpha2}
\end{figure}

Figures~\ref{fig:lineplot_alpha1}--\ref{fig:lineplot_alpha2} illustrate the concordance between Cox (red solid lines) and Fine-Gray (blue dashed lines) estimates across the four correlation levels. At low competing event rates (upper panels showing $\lambda_2 = 0.005, 0.008, 0.01$), the two methods tracked closely together across all values of $\theta_2$, with minimal divergence. As competing rates increased (lower panels showing $\lambda_2 = 0.02, 0.03, 0.05$), the methods began to diverge, particularly when treatment effects on competing events differed substantially from effects on primary events ($\theta_2$ far from 0.80). However, even at the highest competing rate examined (5\% annually), the divergence between Cox and Fine-Gray remained modest when treatment effects were directionally consistent ($\theta_2$ between 0.7 and 1.0).

When competing event rates remained at or below 2\% annually, the range typical of contemporary CVOTs, the absolute difference between Cox and Fine-Gray estimates remained small (typically less than 0.02 on the hazard ratio scale) regardless of treatment effect direction or event correlation, provided treatment effects were not extremely discordant. {As competing rates increased toward the upper end of our simulation range (5\% annually), differences between methods became more apparent but remained modest when $\theta_2$ was close to $\theta_1$. Substantial divergence emerged only at the highest competing rate examined (5\% annually) combined with highly discordant treatment effects ($\theta_2 < 0.7$ or $\theta_2 > 1.0$).} This pattern held consistently across all levels of event correlation examined.

{\subsection{Estimand Gap: Observed Cause-Specific HR versus Latent Treatment Effect}}

{Beyond comparing Cox and Fine-Gray methods with each other, we examined the estimand gap, the difference between what the Cox model estimates (the observed cause-specific HR, $\text{HR}_{\text{CS}}$) and the latent treatment effect parameter $\theta_1$. The closed-form expression in Equation~(4) predicts that $\text{HR}_{\text{CS}} = \theta_1$ when $\alpha = 1$ (independence), $\theta_2 = \theta_1$ (concordant effects), or $\lambda_2 \to 0$ (negligible competing risk). Our simulations confirmed this analytical prediction: the mean Cox HR matched the theoretical $\text{HR}_{\text{CS}}$ from Equation~(4) to within Monte Carlo noise across all 528 base-case scenarios (mean absolute deviation $< 0.01$ on the HR scale for $\lambda_2 \leq 0.05$).}

\begin{figure}[ht]
\centering
\includegraphics[width=\textwidth]{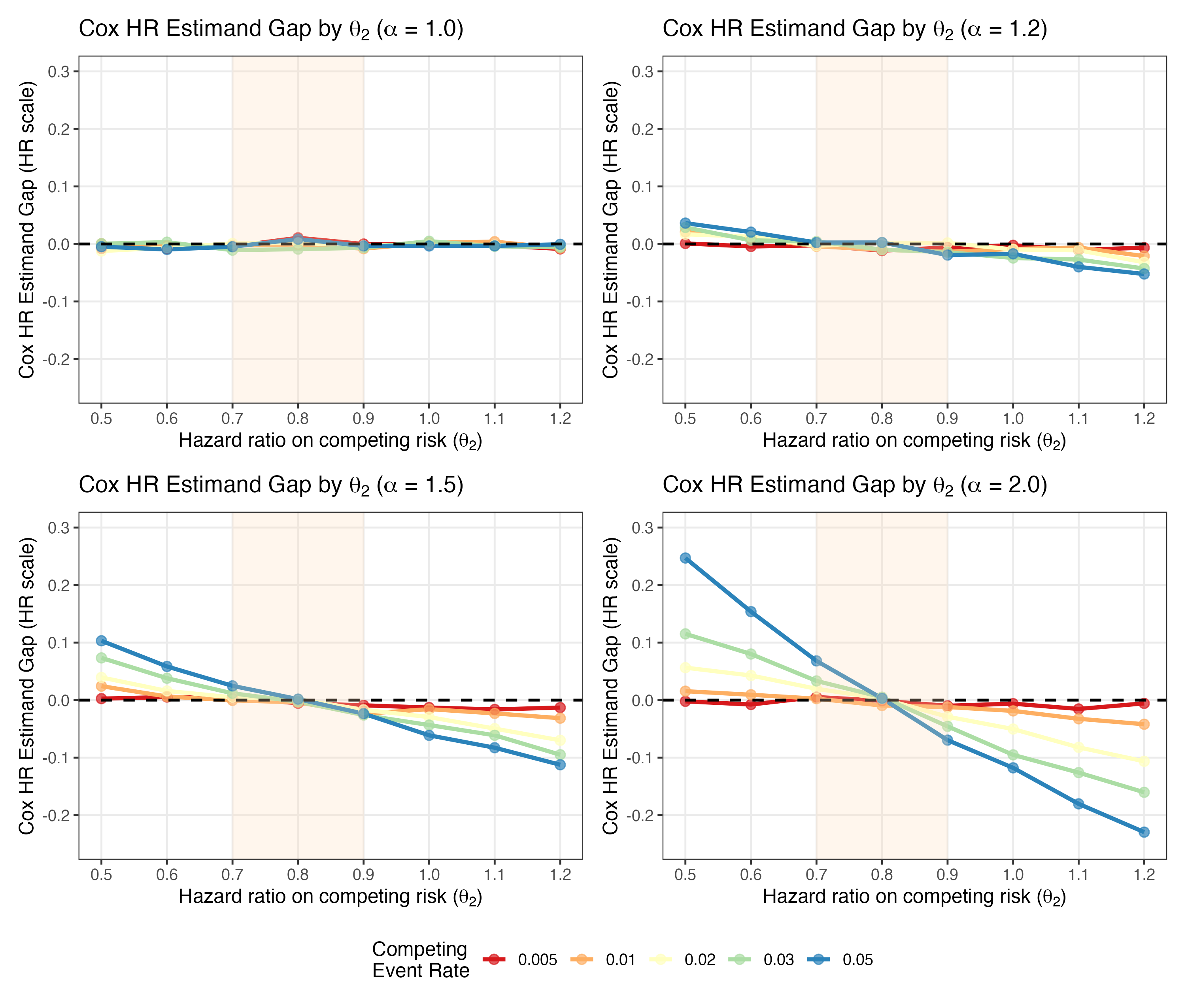}
\caption{{Cox HR estimand gap by $\theta_2$ across correlation levels. The estimand gap ($\widehat{\text{HR}}_{\text{Cox}} - 0.80$) is shown across different treatment effects on the competing risk ($\theta_2$, x-axis) for different competing event rates (colored lines). Four panels represent different correlation strengths ($\alpha = 1.0, 1.2, 1.5, 2.0$). The shaded region ($\theta_2 = 0.7$--$0.9$) demonstrates that the gap vanishes when treatment effects on primary and competing events are concordant, consistent with Equation~(4).}}
\label{fig:bias_valley_combined}
\end{figure}

\begin{figure}[ht]
\centering
\includegraphics[width=\textwidth]{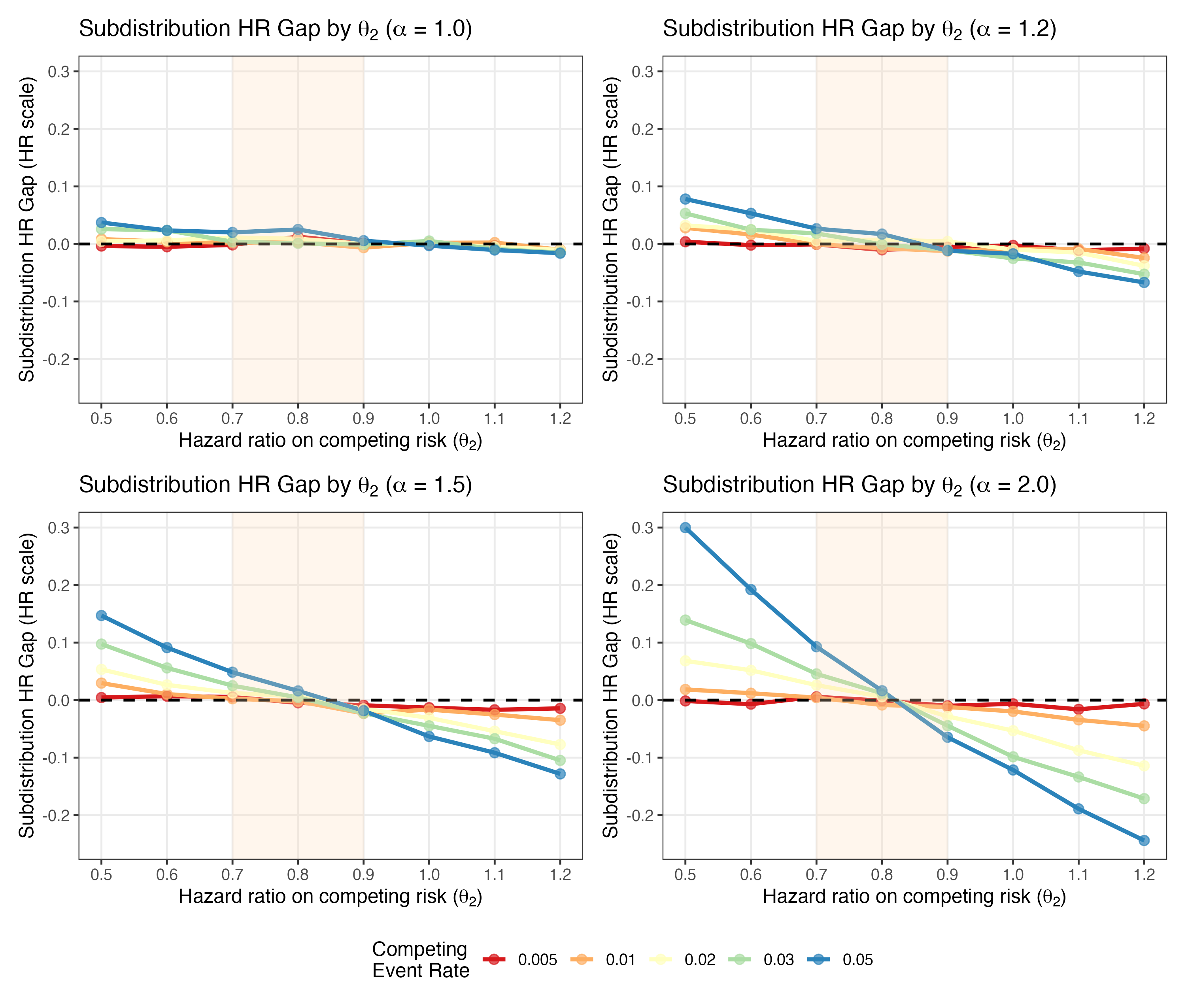}
\caption{{Fine-Gray sHR estimand gap by $\theta_2$ across correlation levels. Format as in Figure~\ref{fig:bias_valley_combined} but showing $\widehat{\text{sHR}}_{\text{FG}} - 0.80$. The Fine-Gray sHR shows a qualitatively similar pattern: the gap vanishes near $\theta_2 = \theta_1$. At high competing rates with $\theta_2 < 0.80$, the sHR gap is wider than the Cox gap (positive direction); with $\theta_2 > 0.80$, the sHR gap is wider in the negative direction.}}
\label{fig:bias_valley_combined_fg}
\end{figure}

{Figures~\ref{fig:bias_valley_combined}--\ref{fig:bias_valley_combined_fg} present the estimand gap ($\widehat{\text{HR}} - 0.80$) for Cox and Fine-Gray respectively across all four correlation levels, demonstrating a critical pattern:} \textbf{{the estimand gap was minimal when the treatment effect on competing events ($\theta_2$) was similar to the treatment effect on primary events ($\theta_1 = 0.80$), regardless of competing event rate or correlation structure.}} {The shaded region ($\theta_2 = 0.7$--$0.9$) highlights this concordance locus, where all curves converge near zero regardless of $\lambda_2$ or $\alpha$. This pattern follows directly from Equation~(4): when $\theta_2 = \theta_1$, the bracketed term equals 1 and $\text{HR}_{\text{CS}} = \theta_1$ exactly.}

{Both estimators provided values close to $\theta_1$ under specific conditions. When primary and competing events were independent ($\alpha = 1.0$), both methods closely approximated 0.80 regardless of competing event rate or $\theta_2$, consistent with Equation~(4) where the exponent $1/\alpha - 1 = 0$, making $\text{HR}_{\text{CS}} = \theta_1$. Similarly, when $\theta_2 = \theta_1 = 0.80$, both methods yielded estimates near 0.80 regardless of correlation or competing rate. Outside these conditions, both estimators deviated from $\theta_1$, with the magnitude determined by the interaction of $\lambda_2$, $\theta_2$, and $\alpha$ as predicted by the closed-form expression.}

{When treatment reduced competing events more strongly than primary events ($\theta_2 < 0.7$), both Cox and Fine-Gray showed positive estimand gaps that increased with competing event rates. Conversely, when treatment had neutral or adverse effects on competing events ($\theta_2 > 0.9$), negative gaps emerged. Across the parameter range typical of CVOTs (competing rate $\leq 2\%$, $\theta_2$ between 0.7 and 1.0), the estimand gap remained minimal for both methods regardless of correlation level. Correlation modulated the gap magnitude but did not alter these patterns, as the $\alpha$ exponent in Equation~(4) amplifies (when $\alpha > 1$) or eliminates (when $\alpha = 1$) the effect of discordant treatment effects.}

\subsection{Robustness to Simulation Assumptions}

To assess whether the findings above depend on specific modeling choices in the primary DGM, we conduct targeted sensitivity analyses varying five assumptions. In each analysis, all other parameters are held at base-case values and a single assumption is relaxed. {All robustness figures (Supplementary Figures S1--S5) overlay both Cox and Fine-Gray estimates on the same axes (solid vs.\ dashed lines) and report Monte Carlo standard errors in the accompanying results tables.}

\textbf{Sample size.} The {estimand gap} patterns reported in Sections 3.1--3.2 were invariant to sample size: simulations at $n = 1{,}000$ and $n = 2{,}000$ showed identical {gap} magnitudes with only improved precision (standard errors scaling as $1/\sqrt{n}$; Figure~S1).

\textbf{Non-proportional hazards.} We replaced the constant treatment effect $\theta_1 = 0.80$ with two piecewise-constant profiles: a lag scenario ($\theta_1(t) = 1.0$ for $t < 1$ year, $0.70$ thereafter) and a waning scenario ($\theta_1(t) = 0.60$ for $t < 2$ years, $1.0$ thereafter). {Under these NPH profiles, the time-averaged Cox HR replaces the constant $\theta_1$ as the reference; we note that a single ``true HR'' is not well-defined when $\theta_1(t)$ varies. Nevertheless, the qualitative pattern of the estimand gap was governed by $\lambda_2$, $\theta_2$, and $\alpha$, not by the temporal shape of the primary treatment effect (Figure~S2).} At clinically relevant rates, all three profiles produced nearly identical estimates.

\textbf{Baseline hazard shape.} We replaced exponential marginals with Weibull distributions: increasing hazard (shape $= 1.5$) and decreasing hazard (shape $= 0.75$). {Scale parameters were recalibrated so that the expected cumulative event \emph{probability} (not merely the instantaneous hazard) over the [3, 5]-year follow-up window matched the exponential base case, isolating the effect of hazard shape from event incidence (see Supplementary Materials). We draw this distinction explicitly because, under a non-constant Weibull hazard, matching the instantaneous annual hazard rate across shapes does not equalize the cumulative event probability over follow-up; it is the latter, not the former, that is clinically and statistically relevant for comparing scenarios.} Under independence ($\alpha = 1$), all three baseline shapes yielded identical results. Under strong dependence with elevated competing rates ($\alpha = 2$, $\lambda_2 = 0.05$), the increasing-hazard Weibull amplified the {gap} magnitude but did not change its direction or the concordance-locus pattern (Figure~S3). The constant-hazard assumption was not driving the primary findings.

\textbf{Dependence structure.} We replaced the Gumbel-Hougaard copula (upper-tail dependence) with Clayton (lower-tail dependence), Frank (symmetric), and Frank with negative association ($\tau = -0.33$), all calibrated to matched Kendall's $\tau$ for comparability. At typical CVOT competing rates ($\lambda_2 = 0.01$), all copula families produced indistinguishable results. At elevated rates ($\lambda_2 = 0.05$), positive copulas at the same $\tau$ produced the same qualitative pattern, with Gumbel giving the widest {gap} range, representing an upper bound among positive-dependence families. The sole qualitatively different finding was under negative dependence, which reversed the direction of the gap (Figure~S4). This scenario requires the biologically unusual assumption that CV and non-CV death are inversely correlated.

\textbf{Patient-level heterogeneity.} We introduced a shared Gamma frailty ($\text{Var}[Z] \in \{0.5, 1.0, 2.0\}$) multiplying both event hazards, representing unobserved patient-level risk variation. Frailty attenuated the observed marginal HR toward 1.0, a well-known selection effect, but preserved the qualitative {estimand gap} pattern (Figure~S5). The magnitude of attenuation remained $\leq 0.03$ on the HR scale, confirming that this assumption is not load-bearing for the primary conclusions.

In summary, the primary findings were robust across all five sensitivity dimensions examined. {All Cox and Fine-Gray model fits converged successfully across every simulation scenario in all five robustness suites, with no model-fitting failures observed.}

\section{Empirical Illustration} \label{sec:empirical}

To complement the simulation findings, we present REWIND \citep{gerstein2019dulaglutide} as a real-data illustration. REWIND randomized 9,901 patients with type 2 diabetes to dulaglutide 1.5 mg or placebo, with a median follow-up of 5.4 years and a non-CV death rate of approximately 1\% annually, a typical CVOT setting. {This illustration analyzes individual patient-level REWIND data. The analysis population was the intention-to-treat population as defined in the original trial \citep{gerstein2019dulaglutide}, with the same censoring rules as the primary REWIND analysis. Cox models were fit using the \texttt{survival} package in R \citep{therneau2023survival}, and Fine-Gray models using the \texttt{cmprsk} package in R \citep{gray2022cmprsk}, consistent with the simulation methodology in Section~\ref{sec:methods}.} Table~\ref{tab:empirical} shows Cox and Fine-Gray results for MACE-3 and its individual components, each with its appropriate competing event.

\begin{table}[htbp]
\centering
\caption{Cox HR and Fine-Gray sHR for MACE-3 and Components in the REWIND Trial}
\label{tab:empirical}
\footnotesize
\begin{tabular*}{\textwidth}{@{\extracolsep{\fill}} l c c c c}
\toprule
\textbf{Endpoint} & \textbf{Events (Dula/PBO)} & \textbf{Cox HR (95\% CI)} & \textbf{FG sHR (95\% CI)} & \textbf{Competing event} \\
\midrule
\textbf{MACE-3 (composite)} & 594 / 663 & 0.88 (0.79--0.98) & 0.89 (0.79--0.99) & Non-CV death \\
\quad CV death & 317 / 346 & 0.91 (0.78--1.06) & 0.91 (0.78--1.06) & Non-CV death \\
\quad Non-fatal MI & 205 / 212 & 0.96 (0.79--1.16) & 0.97 (0.80--1.17) & All death \\
\quad Non-fatal stroke & 135 / 175 & 0.76 (0.61--0.95) & 0.77 (0.61--0.96) & All death \\
\addlinespace
All-cause death & 536 / 592 & 0.90 (0.80--1.01) & NA & --- \\
\bottomrule
\end{tabular*}
\begin{flushleft}
\footnotesize
REWIND: dulaglutide 1.5 mg vs placebo, $N = 9{,}901$, ITT population. Cox HR from cause-specific model (competing events censored). FG sHR from Fine-Gray subdistribution hazard model. Competing events: non-CV death for MACE-3 and CV death; all-cause death for non-fatal MI and stroke. All-cause death has no competing event.
\end{flushleft}
\end{table}

Across all four endpoints where both methods apply, the numerical difference between Cox HR and Fine-Gray sHR was $\leq 0.01$ on the hazard ratio scale. This concordance held regardless of whether the competing event was non-CV death alone (for the composite and CV death) or all-cause death (for non-fatal MI and stroke), and regardless of the magnitude of the treatment effect (ranging from HR 0.76 for stroke to 0.96 for MI). The pattern was precisely what our simulations predicted for a trial with low competing event rates: when non-CV death occurred at $\approx 1\%$ annually, the conditioning sets of the two models were nearly identical, and the estimand distinction had negligible practical impact.

A systematic cross-industry meta-analysis comparing Cox and Fine-Gray results across a broader set of CVOTs would further strengthen this empirical evidence and is an important direction for future collaborative work.

\section{Discussion} \label{sec:discussion}

Across the parameter ranges relevant to CV outcome trials, our simulations and the empirical illustration in Section~\ref{sec:empirical} converge on the same conclusion: Cox and Fine-Gray methods produced concordant results when competing event rates were low. What drove their divergence was not event correlation per se, but the interaction between the competing event rate and the direction of treatment effects on the competing event. For modern therapies with pleiotropic mechanisms (GLP-1 receptor agonists being a leading example), treatment benefits tend to be concordant across event types, and methods remained aligned even at moderately elevated competing rates. {Substantial divergence emerged only when competing rates reached the upper end of our simulation range (5\% annually) combined with highly discordant treatment effects on primary and competing events, a combination uncommon in contemporary CVOTs.}

{The Cox model consistently estimated the observed cause-specific hazard ratio $\text{HR}_{\text{CS}}$, which we showed analytically (Equation~4) to differ from the latent treatment effect parameter $\theta_1$ by an estimand gap that depends on $\lambda_2$, $\theta_2$, and $\alpha$. This gap was negligible when events were independent, when treatment effects were concordant, or when competing rates were low, precisely the conditions prevailing in typical CVOTs.}

\subsection{Dependent vs. Independent Competing Risks}

Not all competing risks warrant special analytical consideration. NEJM statistical guidance distinguishes between \textit{dependent} and \textit{independent} competing risks \citep{nejm2022guidelines}. A competing risk is considered dependent when it is correlated with the event of interest, such that patients experiencing the competing event would have had different risk of the primary outcome compared to those who remain event-free. Our simulation study considers both dependent and independent competing risks by varying the strength of correlation and different treatment effects.

The practical implication is that investigators should consider the likely degree of dependence between competing events and primary events when deciding whether competing risk methods add value. When competing events are truly independent, standard Cox models are appropriate regardless of competing event rates. When competing events are dependent (typical in CVOTs with non-CV death), our simulations showed that {Fine-Gray models yielded nearly identical numerical results to Cox models} as long as competing rates remained low and treatment effects on both event types pointed in the same direction.

\subsection{Cox Model Remains Valid in Typical CVOT Settings}

For the majority of CV outcome trials, our findings support several practical recommendations. {When competing event rates are expected to remain rare compared to the primary endpoint, Cox proportional hazards models treating competing events as censored consistently estimate the cause-specific hazard ratio, and the estimand gap relative to the latent treatment effect is negligible.} Even without formal competing risk modeling, transparently reporting descriptive summaries of competing events by treatment group provides valuable context. {In typical CVOT settings where competing event rates remained low ($\leq 2\%$ annually), we did not expect substantial differences between Cox and Fine-Gray methods, and the cause-specific Cox model remains appropriate for the primary analysis targeting MACE rate reduction.} {When cumulative incidence is the quantity of clinical interest, for example, communicating absolute risk reductions to patients, the Aalen-Johansen estimator or Fine-Gray framework is the appropriate complementary approach.}

\subsection{Importance of Pre-Specified Analysis}
Pre-specified analytical approaches should take priority, and pre-specified Cox models should not be abandoned post-hoc simply because competing risks appear relevant after unblinding data. This principle is fundamental to maintaining trial integrity and is emphasized across multiple regulatory guidelines.

\subsection{Fine-Gray Model is Not a Sensitivity Analysis to Cox Model}

{A key principle highlighted in the ICH E9(R1) Addendum on Estimands and Sensitivity Analysis is that ``sensitivity analyses should investigate the robustness of the overall conclusions to deviations from assumptions made in the primary analysis,'' and should target the same estimand as the primary analysis except for the assumption under evaluation. In many CV outcome trials, however, it has become common practice to conduct Fine-Gray competing risk analyses as ``sensitivity analyses'' to the Cox proportional hazards model. We emphasize that this practice is misaligned with the ICH E9(R1) framework, not because the Fine-Gray model lacks validity, but because it targets a fundamentally different estimand.

These are complementary estimands addressing different scientific questions: the cause-specific HR characterizes treatment effects on the instantaneous event rate among those currently at risk, while the subdistribution HR characterizes treatment effects on cumulative risk in the presence of competing events. Neither is intrinsically superior; the choice should be driven by the scientific question of interest \citep{austin2017practical,andersen2012competing}.

In the context of CVOT primary analyses where the regulatory question concerns treatment efficacy on MACE occurrence, the cause-specific HR (Cox model) aligns naturally with the clinical reasoning: ``Does treatment reduce the rate of MACE among event-free patients?'' When cumulative incidence is the quantity of clinical interest, for example, when communicating absolute risk reductions to patients, the Fine-Gray framework or nonparametric Aalen-Johansen estimation is more appropriate. Comparing Cox and Fine-Gray estimates does not constitute a sensitivity analysis because differences between the two reflect that the models target different estimands, not lack of robustness of either estimand individually. Proper sensitivity analyses for the cause-specific HR should instead modify assumptions relevant to that same estimand (e.g., inverse probability of censoring weighting to assess departures from independent censoring).}

\subsection{Include All-Cause Mortality in Composite Endpoint}

An important alternative that sidesteps competing risk complications entirely is including all-cause mortality in the composite endpoint, particularly when treatment demonstrates benefits for both CV and non-CV death. Clinical trials of incretin-based therapies such as SELECT and SOUL, along with meta-analyses, demonstrate that GLP-1 receptor agonists benefit multiple metabolic risk factors and reduce all-cause death risk \citep{lincoff2023semaglutide,mcguire2025oral,lee2025cardiovascular}. Including all-cause mortality in the composite offers several advantages: it captures the full mortality benefit without artificial partitioning into competing components, it avoids the complexities of competing risk modeling, it eliminates concerns about cause-of-death misclassification, and it aligns with patient-centered outcomes. However, including all-cause death in the composite endpoint changes the clinical question of interest and may dilute the treatment effect if non-CV death is not expected to benefit from treatment.

\subsection{Alternative Methods to Handle Competing Risk}

{The estimand gap between the observed cause-specific HR and the latent treatment effect under high competing-risk scenarios} motivates consideration of alternative approaches when marginal causal effects are the target of inference. Inverse probability of censoring weighting (IPCW) methods model the probability that patients remain uncensored and weight observations by inverse probabilities \citep{austin2025inverse}. {We note, however, that IPCW applied to competing events changes the inferential target rather than merely testing an assumption of the cause-specific hazard approach.} However, in CVOTs with MACE-3 as the primary endpoint, modeling the probability of non-CV death also suffers from competing risk from CV death, creating a circular dependency. Multiple imputation approaches account for the fact that patients who die tend to be at higher risk for the unobserved outcome \citep{gregson2024competing,wang2023missing}. Multistate models explicitly model transitions between health states \citep{cook2018multistate}, providing rich information about disease progression but increasing analytical complexity.

\subsection{Study Limitations}

Despite the robustness checks in Section~\ref{sec:results}, several limitations remain.

\textbf{Semi-competing risks and illness-death structure.} MACE-3 includes non-fatal MI and non-fatal stroke, which are non-terminal events that can precede the terminal event (death). This defines an illness-death (semi-competing risk) structure where a non-terminal event may be followed by death, but death precludes future non-terminal events \citep{putter2007tutorial,andersen2002multi}. Our DGM models only the time to first event from each type, a standard simplification that is defensible for first-MACE analysis but does not capture the non-terminal $\to$ terminal transition. Full illness-death or multistate modeling is beyond the scope of this study.

\textbf{Scope.} Our simulation focuses on the typical CVOT setting where competing event rates are relatively low ($\leq 5\%$ annually). For clinical trials in other therapeutic areas such as oncology or geriatric populations, where competing event rates can be substantially higher and event correlation may follow different patterns, the conclusions in this paper may not directly apply. The robustness analyses, while covering five key dimensions, do not exhaust all possible DGM configurations; for example, we did not examine time-varying dependence structures or informative censoring mechanisms beyond the competing event itself.

\section{Conclusion} \label{sec:conclusion}

The concordance of Cox and Fine-Gray methods in typical CVOT settings (where competing event rates remain low) provides reassurance that analytical approach selection, while important to pre-specify, had limited impact on trial conclusions under these conditions. We derived a closed-form expression for the observed cause-specific hazard ratio under the Gumbel copula and verified that it predicted the simulation results across all 528 base-case scenarios. The Cox model consistently estimated this quantity, and the estimand gap between the observed cause-specific HR and the latent treatment effect parameter $\theta_1$ was negligible across the parameter range typical of contemporary CVOTs. These findings were robust across five sensitivity dimensions: sample size, baseline hazard shape, treatment effect non-proportionality, dependence structure, and patient-level heterogeneity.

The choice between analytical methods should be driven by the scientific estimand. When the cause-specific treatment effect on MACE rate among at-risk subjects is the primary target, Cox proportional hazards models treating competing events as censored are appropriate. When cumulative incidence is the quantity of clinical interest, for example communicating absolute risk reductions to patients, the Fine-Gray model or Aalen-Johansen estimation is the appropriate approach. The choice should be pre-specified according to the estimand of interest, rather than applied post hoc based on observed competing event patterns. Comparing Cox and Fine-Gray results does not constitute a sensitivity analysis under ICH~E9(R1), as the two models target different estimands. The estimand gap between $\text{HR}_{\text{CS}}$ and $\theta_1$ under high competing-risk scenarios motivates consideration of alternative approaches (inverse probability of censoring weighting, multiple imputation, or inclusion of all-cause mortality in primary composite endpoints) when the marginal causal effect is the target of inference.

\section*{Conflicts of Interest}
All authors are employees and minor shareholders of Eli Lilly and Company.

\section*{Funding Statement}
This research received no grant.

\section*{Data Availability Statement}
{All simulation code and results are available from the authors upon request. The REWIND empirical illustration used individual patient-level data from the REWIND trial \citep{gerstein2019dulaglutide}, which are not publicly available.}

\section*{Ethics Statement}

{This research is primarily a simulation study using Monte Carlo methods with publicly available statistical software. The empirical illustration uses individual patient-level data from the REWIND trial, which was conducted with appropriate ethics approval and informed consent as described in the original publication \citep{gerstein2019dulaglutide}. No additional ethics approval was required for this exploratory methodological analysis.}

\clearpage

\appendix

\section*{Supplementary Materials: Robustness Simulation Specifications}

All robustness suites retain the base-case parameters ($n = 500$, $B = 2{,}000$, $\lambda_1 = 0.035$, $\theta_1 = 0.80$, censoring $\sim \text{Uniform}(3,5)$) except for the single assumption being varied. Grid: $\alpha \in \{1.0, 2.0\}$, $\lambda_2 \in \{0.01, 0.05\}$, $\theta_2 \in \{0.5, 0.8, 1.0, 1.2, 1.5\}$.

\paragraph{S1: Sample size.} Replace $n = 500$ with $n \in \{1{,}000,\; 2{,}000\}$. All other parameters unchanged. Seeds: $20250618 + n$.

\paragraph{S2: Non-proportional hazards.} Replace the constant $\theta_1 = 0.80$ with a piecewise-constant hazard ratio for the primary event:
\begin{align*}
\text{Lag profile:} \quad \theta_1(t) &= \begin{cases} 1.00 & t < 1 \text{ year} \\ 0.70 & t \geq 1 \text{ year} \end{cases} \\[6pt]
\text{Wane profile:} \quad \theta_1(t) &= \begin{cases} 0.60 & t < 2 \text{ years} \\ 1.00 & t \geq 2 \text{ years} \end{cases}
\end{align*}
Event times for treated subjects ($Z = 1$) are generated via inversion of the piecewise-exponential cumulative hazard: given $U_1$ from the copula, solve $\Lambda_1(T_1) = -\log(U_1)$ where $\Lambda_1(t) = \int_0^t \lambda_1 \cdot \theta_1(s)\, ds$ is the cumulative hazard with piecewise-constant rate $\lambda_1 \cdot \theta_1(t)$. Control subjects ($Z = 0$) retain rate $\lambda_1$ throughout. The competing event $T_2$ retains the constant $\theta_2$.

\paragraph{S3: Weibull baseline distributions.} Replace exponential marginals with Weibull proportional hazards:
\[
h_j(t \mid Z) = k_j \cdot \lambda_j^{k_j} \cdot t^{k_j - 1} \cdot \theta_j^Z, \quad j = 1, 2
\]
where $k_j$ is the Weibull shape parameter. Inversion from copula uniform $U_j$:
\[
T_j = \frac{\left(-\log U_j \,/\, \theta_j^Z\right)^{1/k_j}}{\lambda_j}
\]
Under this parametrization, $\theta_j$ remains the hazard ratio regardless of $k_j$. When $k_j = 1$, this reduces to the exponential base case. We test $k_j \in \{0.75, 1.0, 1.5\}$ (decreasing, constant, and increasing baseline hazard), applied identically to both event types.

\paragraph{S4: Alternative dependence structures.} Replace the Gumbel-Hougaard copula with alternative bivariate copulas from the \texttt{copula} R package, calibrated to matched Kendall's $\tau$:
\begin{center}
\small
\begin{tabular}{llcl}
\toprule
\textbf{Copula} & \textbf{Tail dependence} & \textbf{$\tau$} & \textbf{Parameter} \\
\midrule
Gumbel & Upper tail & $+0.333$ & $\alpha = 1.5$ \\
Clayton & Lower tail & $+0.333$ & $\alpha = 1.0$ \\
Frank & None (symmetric) & $+0.333$ & $\alpha = 3.30$ \\
Frank (negative) & None (symmetric) & $-0.333$ & $\alpha = -3.30$ \\
Independence & None & $0$ & --- \\
\bottomrule
\end{tabular}
\end{center}
Given copula draws $(U_1, U_2)$, event times are computed identically to the base case: $T_j = -\log(U_j) / (\lambda_j \cdot \theta_j^Z)$.

\paragraph{S5: Patient-level heterogeneity (shared frailty).} Introduce a subject-specific frailty $Z_i \sim \text{Gamma}(1/\sigma^2,\; 1/\sigma^2)$ with $E[Z_i] = 1$ and $\text{Var}[Z_i] = \sigma^2$. The frailty multiplies both event hazards:
\[
T_{ji} = \frac{-\log(U_{ji})}{Z_i \cdot \lambda_j \cdot \theta_j^{A_i}}, \quad j = 1, 2
\]
where $U_{ji}$ are copula draws and $A_i$ is treatment assignment. When $\sigma^2 = 0$ (degenerate $Z_i = 1$), this reduces to the base case. We test $\sigma^2 \in \{0, 0.5, 1.0, 2.0\}$. The frailty induces additional positive correlation between $T_1$ and $T_2$ beyond the copula, and causes marginal attenuation of the observed HR toward 1.0 over time (frailty selection).

\section*{Supplementary Figures}

\begin{figure}[htbp]
\centering
\includegraphics[width=\textwidth]{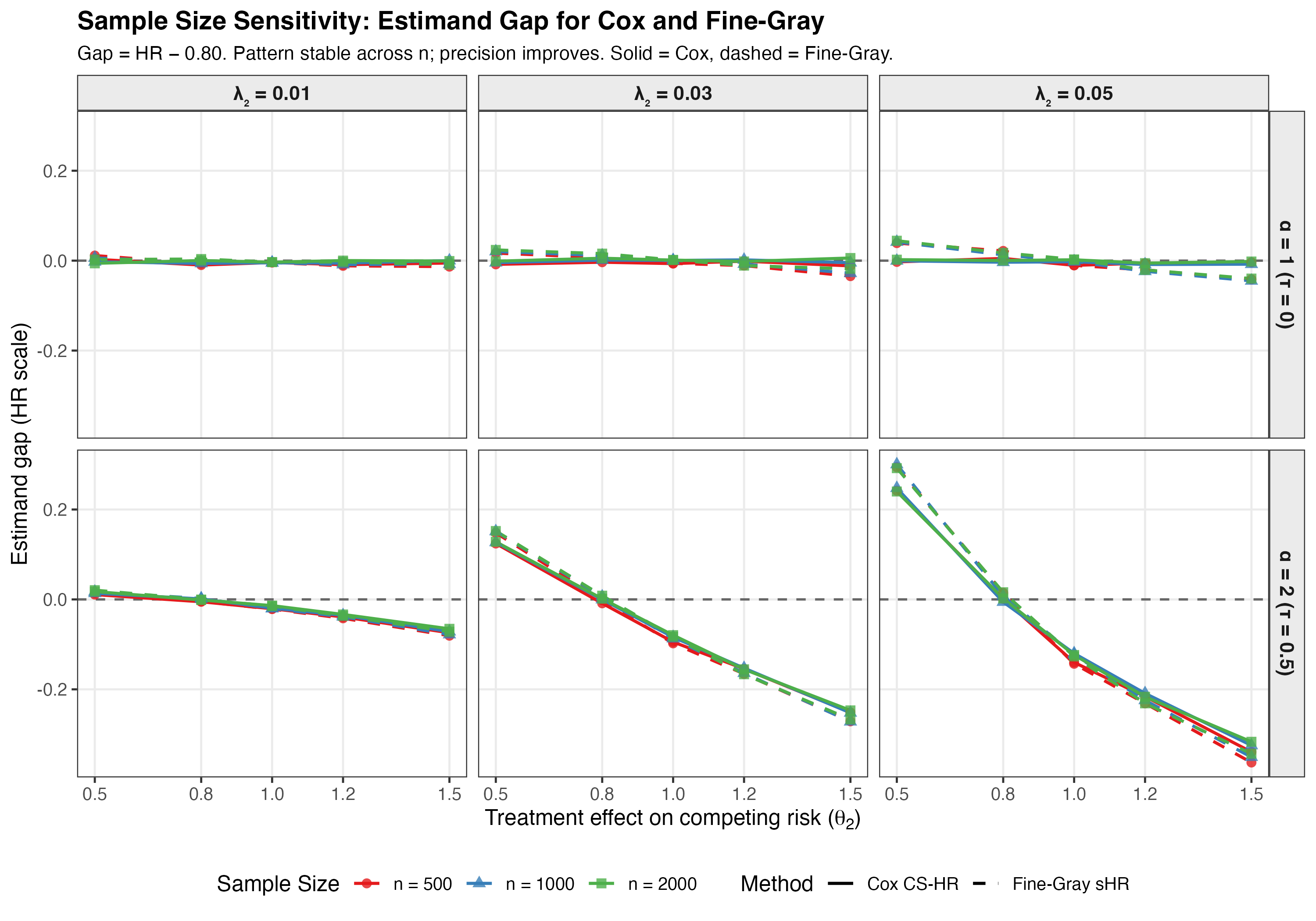}
\caption{{\textbf{Supplementary Figure S1: Sample size sensitivity.} Cox HR (solid lines) and Fine-Gray sHR (dashed lines), both plotted as the estimand gap from $\theta_1$, across sample sizes ($n = 500$, $1{,}000$, $2{,}000$; distinguished by color/line style within each panel). The gap pattern is invariant to sample size; only precision improves. Rows: copula dependence ($\alpha = 1$ independence, $\alpha = 2$ strong). Columns: competing event rate ($\lambda_2$).}}
\label{fig:supp_a}
\end{figure}

\begin{figure}[htbp]
\centering
\includegraphics[width=\textwidth]{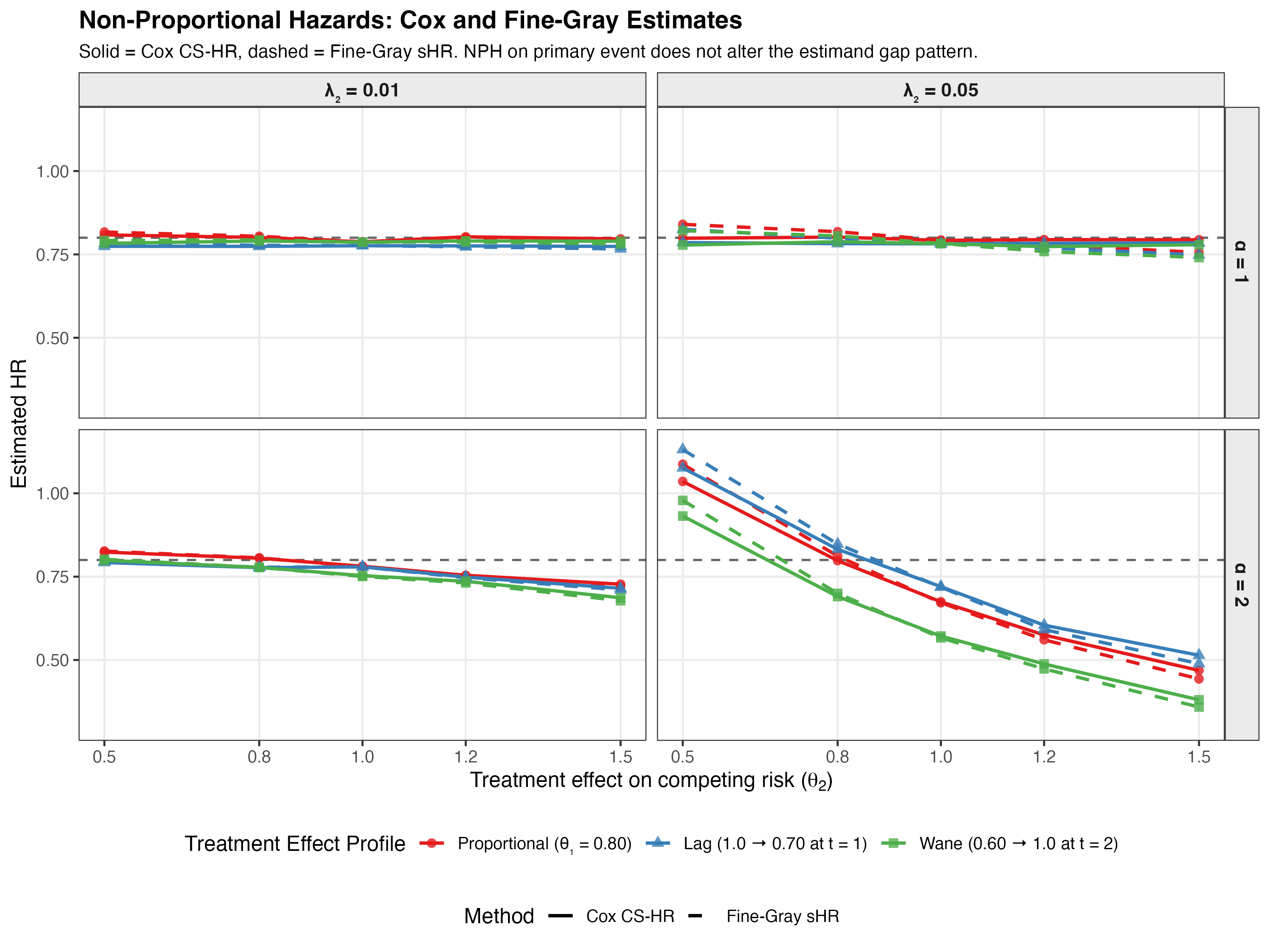}
\caption{{\textbf{Supplementary Figure S2: Non-proportional hazards.} Cox HR and Fine-Gray sHR (overlaid in each panel; solid lines: Cox, dashed lines: Fine-Gray) under proportional (PH, $\theta_1 = 0.80$ constant) and non-proportional treatment effect profiles: Lag ($\theta_1 = 1.0$ for $t < 1$, $0.70$ for $t \geq 1$) and Wane ($\theta_1 = 0.60$ for $t < 2$, $1.0$ for $t \geq 2$). Non-proportional effects on the primary event do not alter the estimand gap pattern. Rows: copula dependence. Columns: competing event rate. Note: under NPH, a single ``true HR'' is not well-defined; the reference is the time-averaged Cox HR.}}
\label{fig:supp_b}
\end{figure}

\begin{figure}[htbp]
\centering
\includegraphics[width=\textwidth]{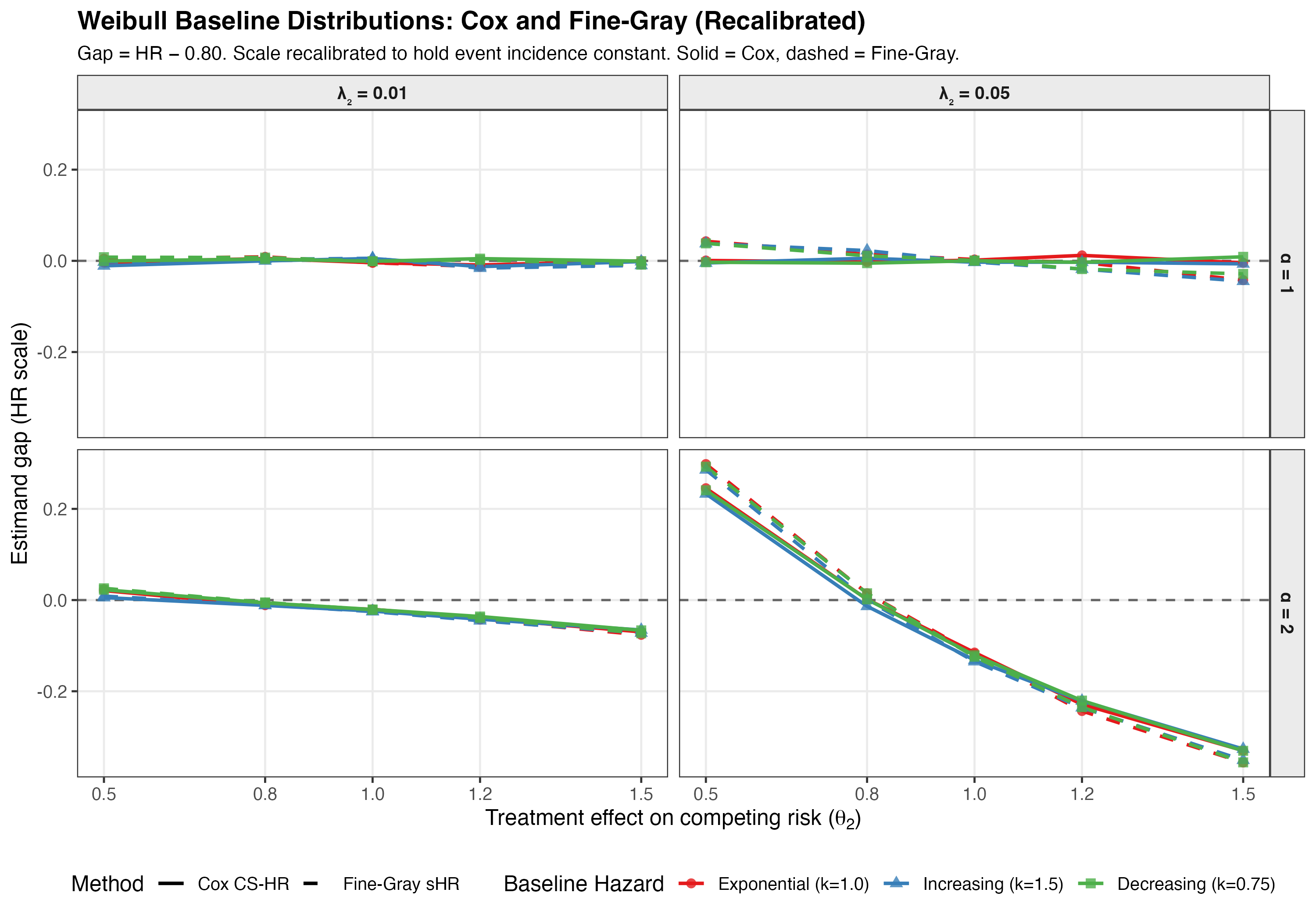}
\caption{{\textbf{Supplementary Figure S3: Weibull baseline distributions (recalibrated).} Cox HR and Fine-Gray sHR (overlaid in each panel; solid lines: Cox, dashed lines: Fine-Gray) under exponential ($k = 1.0$), increasing hazard ($k = 1.5$), and decreasing hazard ($k = 0.75$) marginals; scale parameters recalibrated to hold expected cumulative event probability constant across shapes. Under independence ($\alpha = 1$, top row), all shapes produce identical results. Under strong dependence ($\alpha = 2$), slight magnitude differences emerge but direction is unchanged.}}
\label{fig:supp_c}
\end{figure}

\begin{figure}[htbp]
\centering
\includegraphics[width=\textwidth]{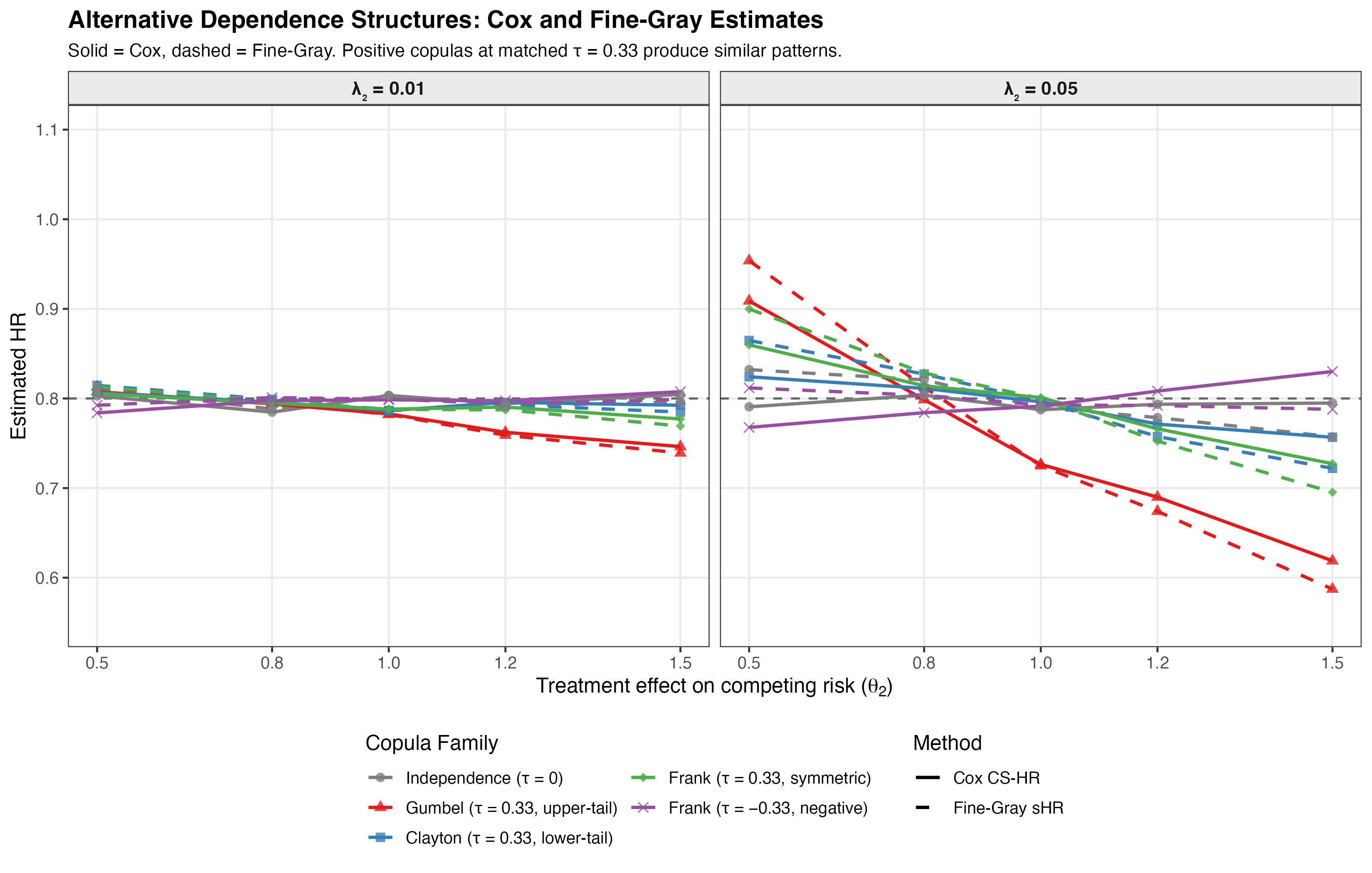}
\caption{{\textbf{Supplementary Figure S4: Alternative copulas.} Cox HR and Fine-Gray sHR (overlaid in each panel; solid lines: Cox, dashed lines: Fine-Gray) under Gumbel (upper-tail), Clayton (lower-tail), Frank (symmetric), and Frank negative ($\tau = -0.33$) copulas, all at matched Kendall's $\tau = 0.33$ for positive families. Left panels ($\lambda_2 = 0.01$): all copulas are indistinguishable. Right panels ($\lambda_2 = 0.05$): positive copulas produce similar patterns; Gumbel is the most extreme. Negative dependence reverses the direction of the estimand gap.}}
\label{fig:supp_d}
\end{figure}

\begin{figure}[htbp]
\centering
\includegraphics[width=\textwidth]{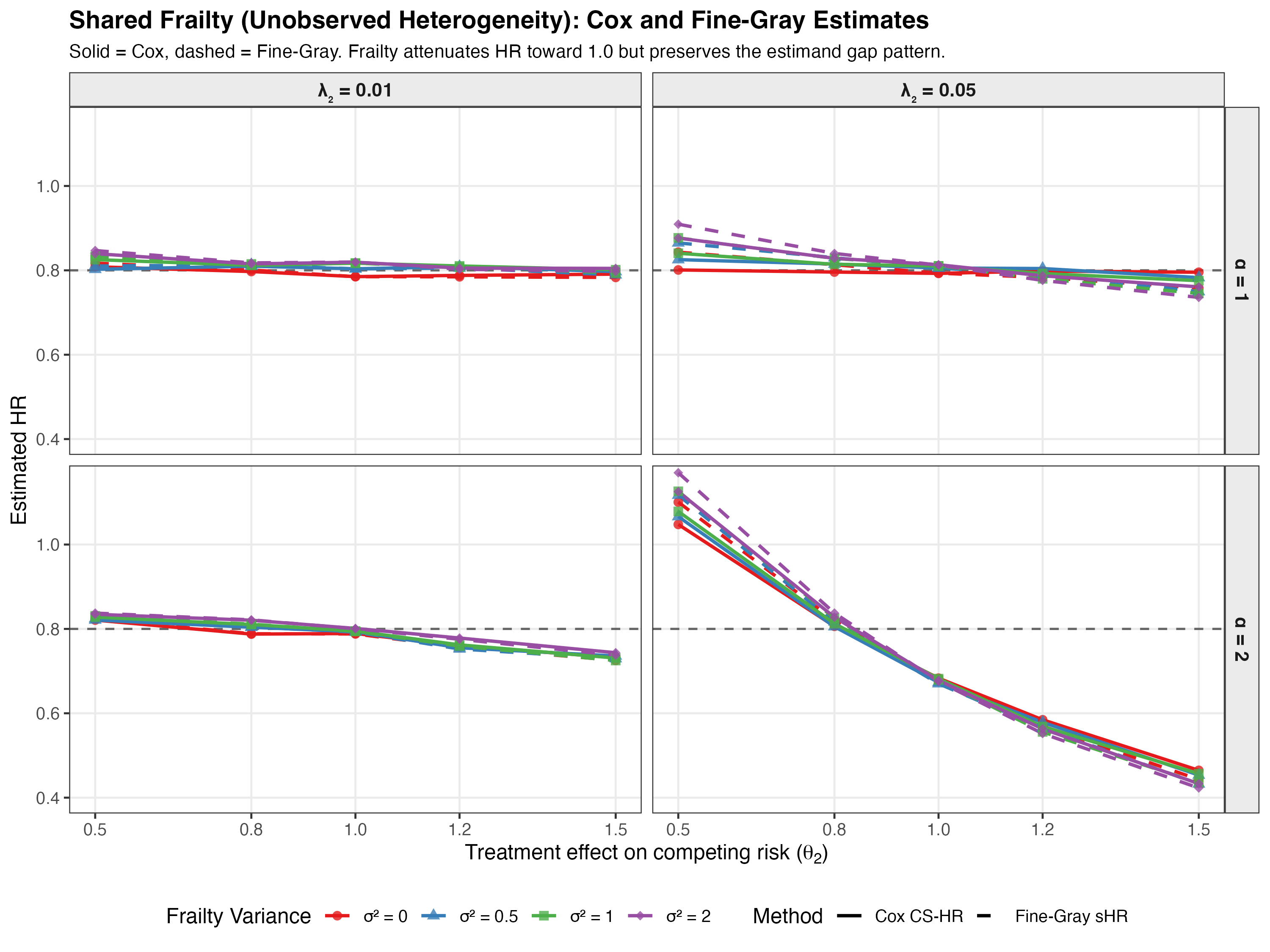}
\caption{{\textbf{Supplementary Figure S5: Shared frailty / heterogeneity.} Cox HR and Fine-Gray sHR (overlaid in each panel; solid lines: Cox, dashed lines: Fine-Gray) under shared Gamma frailty with variance $\sigma^2 \in \{0, 0.5, 1.0, 2.0\}$. Frailty attenuates the marginal HR toward 1.0 (upward shift) but preserves the qualitative estimand gap pattern. Under typical CVOT conditions (top panels), the effect is negligible.}}
\label{fig:supp_e}
\end{figure}

\section*{Appendix A: Derivation of the Observed Cause-Specific Hazard Ratio}

Under the Gumbel-Hougaard copula (Equation~1), the joint survival function evaluated along the diagonal $t_1 = t_2 = t$ (i.e., the probability that a subject is free of \textit{all} events at time $t$) is:
\[
S(t, t \mid Z) = \exp\bigl\{-\bigl[(\lambda_1 \theta_1^Z t)^\alpha + (\lambda_2 \theta_2^Z t)^\alpha\bigr]^{1/\alpha}\bigr\}
\]
Let $A(t_1, t_2 \mid Z) = [(\lambda_1 \theta_1^Z t_1)^\alpha + (\lambda_2 \theta_2^Z t_2)^\alpha]^{1/\alpha}$, so $S = e^{-A}$. The observed cause-specific hazard for event~1 is:
\[
h_1^{\text{cs}}(t \mid Z) = -\frac{\partial}{\partial t_1} \log S(t_1, t_2 \mid Z)\bigg|_{t_1=t_2=t} = \frac{\partial A}{\partial t_1}\bigg|_{t_1=t_2=t}
\]
Computing the partial derivative:
\begin{align*}
\frac{\partial A}{\partial t_1} &= \frac{1}{\alpha}\bigl[(\lambda_1 \theta_1^Z t_1)^\alpha + (\lambda_2 \theta_2^Z t_2)^\alpha\bigr]^{1/\alpha - 1} \cdot \alpha (\lambda_1 \theta_1^Z)^\alpha t_1^{\alpha - 1} \\
&= (\lambda_1 \theta_1^Z)^\alpha \cdot t_1^{\alpha - 1} \cdot \bigl[(\lambda_1 \theta_1^Z t_1)^\alpha + (\lambda_2 \theta_2^Z t_2)^\alpha\bigr]^{1/\alpha - 1}
\end{align*}
Evaluating at $t_1 = t_2 = t$:
\begin{align*}
h_1^{\text{cs}}(t \mid Z) &= (\lambda_1 \theta_1^Z)^\alpha \cdot t^{\alpha - 1} \cdot \bigl[t^\alpha\bigl((\lambda_1 \theta_1^Z)^\alpha + (\lambda_2 \theta_2^Z)^\alpha\bigr)\bigr]^{1/\alpha - 1} \\
&= (\lambda_1 \theta_1^Z)^\alpha \cdot t^{\alpha - 1} \cdot t^{1 - \alpha} \cdot \bigl[(\lambda_1 \theta_1^Z)^\alpha + (\lambda_2 \theta_2^Z)^\alpha\bigr]^{1/\alpha - 1} \\
&= (\lambda_1 \theta_1^Z)^\alpha \cdot \bigl[(\lambda_1 \theta_1^Z)^\alpha + (\lambda_2 \theta_2^Z)^\alpha\bigr]^{1/\alpha - 1}
\end{align*}
The time $t$ cancels because both marginals are exponential (constant hazard). Thus $h_1^{\text{cs}}$ is constant over time and the cause-specific hazard ratio for event~1 is:
\begin{align*}
\text{HR}_{\text{CS}} &= \frac{h_1^{\text{cs}}(t \mid Z=1)}{h_1^{\text{cs}}(t \mid Z=0)} = \frac{(\lambda_1 \theta_1)^\alpha \bigl[(\lambda_1 \theta_1)^\alpha + (\lambda_2 \theta_2)^\alpha\bigr]^{1/\alpha-1}}{\lambda_1^\alpha \bigl[\lambda_1^\alpha + \lambda_2^\alpha\bigr]^{1/\alpha - 1}} \\[4pt]
&= \theta_1^\alpha \cdot \left[\frac{(\lambda_1 \theta_1)^\alpha + (\lambda_2 \theta_2)^\alpha}{\lambda_1^\alpha + \lambda_2^\alpha}\right]^{1/\alpha - 1}
\end{align*}
\textbf{Special cases:}
\begin{itemize}
\item $\alpha = 1$ (independence): the exponent $1/\alpha - 1 = 0$, so $\text{HR}_{\text{CS}} = \theta_1$. Under independence, Cox recovers the latent treatment effect exactly.
\item $\theta_2 = \theta_1$: the ratio in brackets reduces to $\theta_1^\alpha \cdot [\lambda_1^\alpha + \lambda_2^\alpha] / [\lambda_1^\alpha + \lambda_2^\alpha] = \theta_1^\alpha$, and $\theta_1^\alpha \cdot \theta_1^{\alpha(1/\alpha - 1)} = \theta_1$. When treatment effects are concordant, Cox again recovers $\theta_1$.
\item $\lambda_2 \to 0$ (no competing risk): the ratio in brackets $\to 1$, so $\text{HR}_{\text{CS}} \to \theta_1$.
\end{itemize}
These three special cases are confirmed by our simulations in Section~\ref{sec:results}).

\clearpage
\bibliographystyle{apalike}
\bibliography{ms}

@article{cox1972regression,
  author = {Cox, David R},
  title = {Regression models and life-tables},
  journal = {Journal of the Royal Statistical Society: Series B (Methodological)},
  year = {1972},
  volume = {34},
  number = {2},
  pages = {187--220}
}

@article{fine1999proportional,
  author = {Fine, Jason P and Gray, Robert J},
  title = {A proportional hazards model for the subdistribution of a competing risk},
  journal = {Journal of the American Statistical Association},
  year = {1999},
  volume = {94},
  number = {446},
  pages = {496--509}
}

@article{austin2017practical,
  author = {Austin, Peter C and Fine, Jason P},
  title = {Practical recommendations for reporting {F}ine-{G}ray model analyses for competing risk data},
  journal = {Statistics in Medicine},
  year = {2017},
  volume = {36},
  number = {27},
  pages = {4391--4400}
}

@article{andersen2012competing,
  author = {Andersen, Per Kragh and Geskus, Ronald B and de Witte, Theo and Putter, Hein},
  title = {Competing risks in epidemiology: possibilities and pitfalls},
  journal = {International Journal of Epidemiology},
  year = {2012},
  volume = {41},
  number = {3},
  pages = {861--870}
}

@article{young2020causal,
  author = {Young, Jessica G and Stensrud, Mats J and Tchetgen Tchetgen, Eric J and Hern{\'a}n, Miguel A},
  title = {A causal framework for classical statistical estimands in failure-time settings with competing events},
  journal = {Statistics in Medicine},
  year = {2020},
  volume = {39},
  number = {8},
  pages = {1199--1236}
}

@article{austin2025inverse,
  title={Inverse probability of treatment weighting using the propensity score with competing risks in survival analysis},
  author={Austin, Peter C and Fine, Jason P},
  journal={Statistics in Medicine},
  volume={44},
  number={5},
  pages={e70009},
  year={2025},
  publisher={Wiley Online Library}
}

@manual{gray2022cmprsk,
  author = {Gray, Bob},
  title = {cmprsk: Subdistribution Analysis of Competing Risks},
  year = {2022},
  note = {R package version 2.2-11},
  url = {https://CRAN.R-project.org/package=cmprsk}
}

@misc{nejm2022guidelines,
  author = {{New England Journal of Medicine}},
  title = {Statistical Reporting Guidelines: Considerations in Time-to-Event Analyses},
  year = {2022},
  howpublished = {Available at: \url{https://www.nejm.org/author-center/statistical-reporting-guidelines}},
  note = {Accessed November 2024}
}

@article{oakes1989bivariate,
  title={Bivariate survival models induced by frailties},
  author={Oakes, David},
  journal={Journal of the American Statistical Association},
  volume={84},
  number={406},
  pages={487--493},
  year={1989},
  publisher={Taylor \& Francis}
}

@Manual{gumbel2024package,
    title = {gumbel: package for Gumbel copula},
    author = {Christophe Dutang},
    year = {2024},
    note = {R package version 1.10-3},
    url = {https://CRAN.R-project.org/package=gumbel}
}

@article{gregson2024competing,
  title={Competing Risks in Clinical Trials: Do They Matter and How Should We Account for Them?},
  author={Gregson, John and Pocock, Stuart J and Anker, Stefan D and Bhatt, Deepak L and Packer, Milton and Stone, Gregg W and Zeller, Cordula},
  journal={Journal of the American College of Cardiology},
  volume={84},
  number={11},
  pages={1025--1037},
  year={2024},
  publisher={American College of Cardiology Foundation Washington DC}
}

@article{wang2023missing,
  title={Missing data imputation for a multivariate outcome of mixed variable types},
  author={Wang, Tuo and Zilinskas, Rachel and Li, Ying and Qu, Yongming},
  journal={Statistics in Biopharmaceutical Research},
  volume={15},
  number={4},
  pages={826--837},
  year={2023},
  publisher={Taylor \& Francis}
}

@article{mcguire2025oral,
  title={Oral semaglutide and cardiovascular outcomes in high-risk type 2 diabetes},
  author={McGuire, Darren K and Marx, Nikolaus and Mulvagh, Sharon L and Deanfield, John E and Inzucchi, Silvio E and Pop-Busui, Rodica and Mann, Johannes FE and Emerson, Scott S and Poulter, Neil R and Engelmann, Mads DM and others},
  journal={New England Journal of Medicine},
  volume={392},
  number={20},
  pages={2001--2012},
  year={2025},
  publisher={Mass Medical Soc}
}

@article{lincoff2023semaglutide,
  title={Semaglutide and cardiovascular outcomes in obesity without diabetes},
  author={Lincoff, A Michael and Brown-Frandsen, Kirstine and Colhoun, Helen M and Deanfield, John and Emerson, Scott S and Esbjerg, Sille and Hardt-Lindberg, S{\o}ren and Hovingh, G Kees and Kahn, Steven E and Kushner, Robert F and others},
  journal={New England Journal of Medicine},
  volume={389},
  number={24},
  pages={2221--2232},
  year={2023},
  publisher={Mass Medical Soc}
}

@article{gerstein2019dulaglutide,
  title={Dulaglutide and cardiovascular outcomes in type 2 diabetes (REWIND): a double-blind, randomised placebo-controlled trial},
  author={Gerstein, Hertzel C and Colhoun, Helen M and Dagenais, Gilles R and Diaz, Rafael and Lakshmanan, Mark and Pais, Prem and Probstfield, Jeffrey and Riesmeyer, Jeffrey S and Riddle, Matthew C and Ryd{\'e}n, Lars and others},
  journal={The Lancet},
  volume={394},
  number={10193},
  pages={121--130},
  year={2019},
  publisher={Elsevier}
}

@article{perkovic2024effects,
  title={Effects of semaglutide on chronic kidney disease in patients with type 2 diabetes},
  author={Perkovic, Vlado and Tuttle, Katherine R and Rossing, Peter and Mahaffey, Kenneth W and Mann, Johannes FE and Bakris, George and Baeres, Florian MM and Idorn, Thomas and Bosch-Traberg, Heidrun and Lausvig, Nanna Leonora and others},
  journal={New England Journal of Medicine},
  volume={391},
  number={2},
  pages={109--121},
  year={2024},
  publisher={Mass Medical Soc}
}

@article{pitt2021cardiovascular,
  title={Cardiovascular events with finerenone in kidney disease and type 2 diabetes},
  author={Pitt, Bertram and Filippatos, Gerasimos and Agarwal, Rajiv and Anker, Stefan D and Bakris, George L and Rossing, Peter and Joseph, Amer and Kolkhof, Peter and Nowack, Christina and Schloemer, Patrick and others},
  journal={New England Journal of Medicine},
  volume={385},
  number={24},
  pages={2252--2263},
  year={2021},
  publisher={Mass Medical Soc}
}

@book{cook2018multistate,
  title={Multistate models for the analysis of life history data},
  author={Cook, Richard J and Lawless, Jerald F},
  year={2018},
  publisher={Chapman and Hall/CRC}
}

@article{armbruster2024pitfalls,
  title={Pitfalls of choosing a study end point including cardiovascular death in comparative clinical trials},
  author={Armbruster, Stephanie and Skali, Hicham and Wei, Lee-Jen},
  journal={Circulation},
  volume={150},
  number={23},
  pages={1823--1825},
  year={2024},
  publisher={Lippincott Williams \& Wilkins Hagerstown, MD}
}

@article{austin2021fine,
  title={Fine-Gray subdistribution hazard models to simultaneously estimate the absolute risk of different event types: cumulative total failure probability may exceed 1},
  author={Austin, Peter C and Steyerberg, Ewout W and Putter, Hein},
  journal={Statistics in Medicine},
  volume={40},
  number={19},
  pages={4200--4212},
  year={2021},
  publisher={Wiley Online Library}
}

@article{lee2025cardiovascular,
  title={Cardiovascular and kidney outcomes and mortality with long-acting injectable and oral glucagon-like peptide 1 receptor agonists in individuals with type 2 diabetes: a systematic review and meta-analysis of randomized trials},
  author={Lee, Matthew MY and Sattar, Naveed and Pop-Busui, Rodica and Deanfield, John and Emerson, Scott S and Inzucchi, Silvio E and Mann, Johannes FE and Marx, Nikolaus and Mulvagh, Sharon L and Poulter, Neil R and others},
  journal={Diabetes care},
  volume={48},
  number={5},
  pages={846--859},
  year={2025},
  publisher={American Diabetes Association}
}

@article{putter2007tutorial,
  title={Tutorial in biostatistics: competing risks and multi-state models},
  author={Putter, Hein and Fiocco, Marta and Geskus, Ronald B},
  journal={Statistics in Medicine},
  volume={26},
  number={11},
  pages={2389--2430},
  year={2007},
  publisher={Wiley Online Library}
}

@article{andersen2002multi,
  title={Multi-state models for event history analysis},
  author={Andersen, Per Kragh and Keiding, Niels},
  journal={Statistical Methods in Medical Research},
  volume={11},
  number={2},
  pages={91--115},
  year={2002},
  publisher={Sage Publications}
}

@article{beyersmann2009simulating,
  title={Simulating competing risks data in survival analysis},
  author={Beyersmann, Jan and Latouche, Aur{\'e}lien and Buchholz, Anika and Schumacher, Martin},
  journal={Statistics in Medicine},
  volume={28},
  number={6},
  pages={956--971},
  year={2009},
  publisher={Wiley Online Library}
}

@article{latouche2007misspecified,
  title={Misspecified regression model for the subdistribution hazard of a competing risk},
  author={Latouche, Aur{\'e}lien and Boisson, Val{\'e}rie and Chevret, Sylvie and Porcher, Rapha{\"e}l},
  journal={Statistics in Medicine},
  volume={26},
  number={5},
  pages={965--974},
  year={2007},
  publisher={Wiley Online Library}
}

@Manual{therneau2023survival,
    title = {A Package for Survival Analysis in R},
    author = {Terry M Therneau},
    year = {2024},
    note = {R package version 3.7-0},
    url = {https://CRAN.R-project.org/package=survival},
}

@article{tsiatis1975nonidentifiability,
  title={A nonidentifiability aspect of the problem of competing risks},
  author={Tsiatis, Anastasios},
  journal={Proceedings of the National Academy of Sciences},
  volume={72},
  number={1},
  pages={20--22},
  year={1975},
  publisher={National Acad Sciences}
}

\end{document}